\documentclass{article}

\usepackage[english]{babel}

\usepackage[letterpaper,top=2cm,bottom=2cm,left=3cm,right=3cm,marginparwidth=1.75cm]{geometry}

\usepackage{amsmath}
\usepackage{graphicx}
\usepackage{amssymb}
\usepackage[colorlinks=true, allcolors=blue]{hyperref}
\usepackage[algo2e, ruled,vlined,linesnumbered]{algorithm2e}
\usepackage{subcaption}
\usepackage{xcolor,colortbl}
\usepackage{makecell}
\usepackage{hyperref}

\usepackage{appendix}
\usepackage{caption}
\usepackage{multirow}
\usepackage{authblk}
\usepackage{appendix}
\usepackage{caption}

\title{Biological Random Walks: multi-omics integration for disease gene prioritization}

\author[1]{Michele Gentili}
\author[1]{Leonardo Martini}
\author[2]{ Marialuisa Sponziello}
\author[1]{Luca Becchetti}

\affil[1]{Department of Computer, Control, and Management Engineering Antonio Ruberti, Sapienza University of Rome, Rome, Italy}
\affil[2]{Translational and Precision Medicine Department Sapienza University of Rome, Rome, Italy}
\date{}
\newcommand{\pv}{\ensuremath{\mathbf{q}}} %% Personalization vector
\newcommand{\weight}{\ensuremath{\theta}} %% A weight assigned to a gene
\newcommand{\ann}{\ensuremath{\mathcal{A}}} %% Set of annotations - genes
\newcommand{\source}{\ensuremath{\mathcal{F}}} %% Set of annotations - sources
\newcommand{\neigh}{\ensuremath{N}} %% Set of neighbours
\newcommand{\tmat}{\ensuremath{W}} %% Transition matrix
\newcommand{\dsi}{\ensuremath{DSI}} %% Disease specific interaction function
\begin{document}
\maketitle

\begin{abstract}
\textbf{Motivation:} Over the past decade, network-based approaches have proven useful in identifying disease 
modules within the human interactome, often providing insights into key mechanisms 
and guiding the quest for therapeutic targets. 
This is all the more important, since experimental investigation of potential gene candidates is an 
expensive task, thus not always a feasible option.
On the other hand, many sources of biological information exist beyond the interactome and 
an important research direction is the design of effective techniques for their
integration.\\
\textbf{Results:} In this work, we introduce the Biological Random 
Walks (BRW) approach for disease gene prioritization in the human interactome. The proposed framework leverages 
multiple biological sources within an integrated framework. We perform 
an extensive, comparative study of BRW's performance against 
well-established baselines.\\
\textbf{Availability and implementation:} All code is publicly 
available and can be downloaded at 
\url{https://github.com/LeoM93/BiologicalRandomWalks}. We used publicly 
available datasets, details on their retrieval and preprocessing are 
provided in the supplementary material.\\
\textbf{Supplementary material:} Supplementary material available.\end{abstract}

\section{Introduction}
In recent years, through the advent of big data, genomics, and 
quantitative {\it in silico} methodologies, medicine is witnessing 
tremendous advancements towards the understanding of the human 
pathophysiology. \cite{silverman2020molecular}. Gene-disease 
associations have been identified by genome-wide association studies 
(GWAS) \cite{doi:10.1056/NEJMra0808700} and more recently by whole 
exome or whole genome sequencing studies \cite{KOBOLDT201327}. While 
many of the mechanisms underlying these associations remain largely 
unclear, a growing body of research highlights associations between 
groups of interacting proteins and diseases within the so-called 
``human interactome'', representing the cellular network of all 
physical molecular interactions \cite{barabasi2011network}. A key 
feature is that disease proteins do not appear to be uniformly 
scattered across the interactome \cite{menche2015uncovering}, but they 
are prone to participation in common biological activities such as, for 
example, genome maintenance, cell differentiation or growth signaling, 
which are the most relevant pathways in carcinogenesis 
\cite{ozturk2018emerging}. For these reasons, while traditional single 
protein (i.e., magic bullet) approaches have limited effectiveness in 
addressing complex diseases, network-based ones can prove useful in 
identifying disease modules within the interactome, hopefully providing 
insights into key mechanisms and guiding the quest for therapeutic 
targets. Moreover, experimental investigation of potential gene 
candidates is an expensive task, thus not always a feasible option.

The human interactome refers to all protein-protein interactions within 
a cell, including regulatory interaction of transcription factors, 
metabolic enzyme-coupled interactions, protein complexes, and 
kinase/substrate interactions. This network is largely incomplete. 
Currently, more than 140,000 interactions involving over 13,000 
proteins are known (e.g., see 
\cite{korcsmaros2017next,gustafsson2014modules}). The 
interactome-based approach to network medicine 
\cite{barabasi2011network} proved effective for a number of diseases, 
e.g., by identifying putative biomarkers and subtypes, thus allowing a 
principled approach to drug targeting \cite{barabasi2011network, 
ozturk2018emerging}. The need for new disease genes (or disease 
proteins) as putative candidates for diagnosis, prognosis or treatment, 
motivated the development of a number of algorithms for disease genes 
and modules prediction \cite{ghiassian2015disease}.

Two related classes of methodologies have emerged over the last 
decade as the most promising: module-based 
\cite{ghiassian2015disease,barabasi2011network} and network 
propagation \cite{cowen2017network,kohler2008walking} algorithms. 
Module-based algorithms find topological, functional or disease modules 
in the interactome network, on the hypothesis that these represent 
cellular components likely involved in the same disease. Network 
propagation (or diffusion-based) algorithms leverage the information 
flow through nearby proteins in the network from initial (known) 
disease genes as their main ingredient.

While important and effective in many cases, the interactome is only 
one of many sources of biological information. An important research 
direction is the quest for effective techniques that allow the seamless 
integration of rich and heterogeneous biological sources into methods 
that were originally designed to leverage the topological features of 
the interactome 
\cite{dimitrakopoulos2018network,shang2020network,de2007kernel}.

\paragraph{Our contribution} In this work, we introduce the 
Biological Random Walks (BRW) framework for disease gene 
prioritization. The proposed framework leverages the 
integration of multiple biological sources within a propagation-based 
approach. We compare BRW's performance against well-established 
baselines, such as: RWR \cite{Navlakha}, DIAMOnD 
\cite{ghiassian2015disease}, DADA \cite{erten2011d} and RWR-M 
\cite{valdeolivas2019random}. In particular, we investigate BRW's 
performance along different axes: i) an in-depth, comparative analysis 
on four cancer phenotypes (i.e., breast cancer, lung 
adenocarcinoma, papillary thyroid cancer, colorectal adenocarcinoma); 
ii) a broad comparative analysis of BRW's prioritization performance 
over a wide spectrum of Mendelian diseases with different 
characteristics and prior information available; iii) an 
external validation using FDA approved drugs for Breast Cancer 
treatment, along with an evaluation of the algorithm results' stability 
across multiple population studies. \footnote{Some of the ideas presented 
in this submission, albeit in preliminary form and accompanied by a 
minimal, internal validation, appeared in 
\cite{gentili2019biological}. For the sake of completeness, 
Supplementary Section \ref{subse:new_vs_old} presents a comparison of 
our approach with \cite{gentili2019biological}}.

\section{Materials and Methods}
\label{section:materials_methods}

\begin{figure*}[t]
    \centering
    \includegraphics[width=\textwidth]{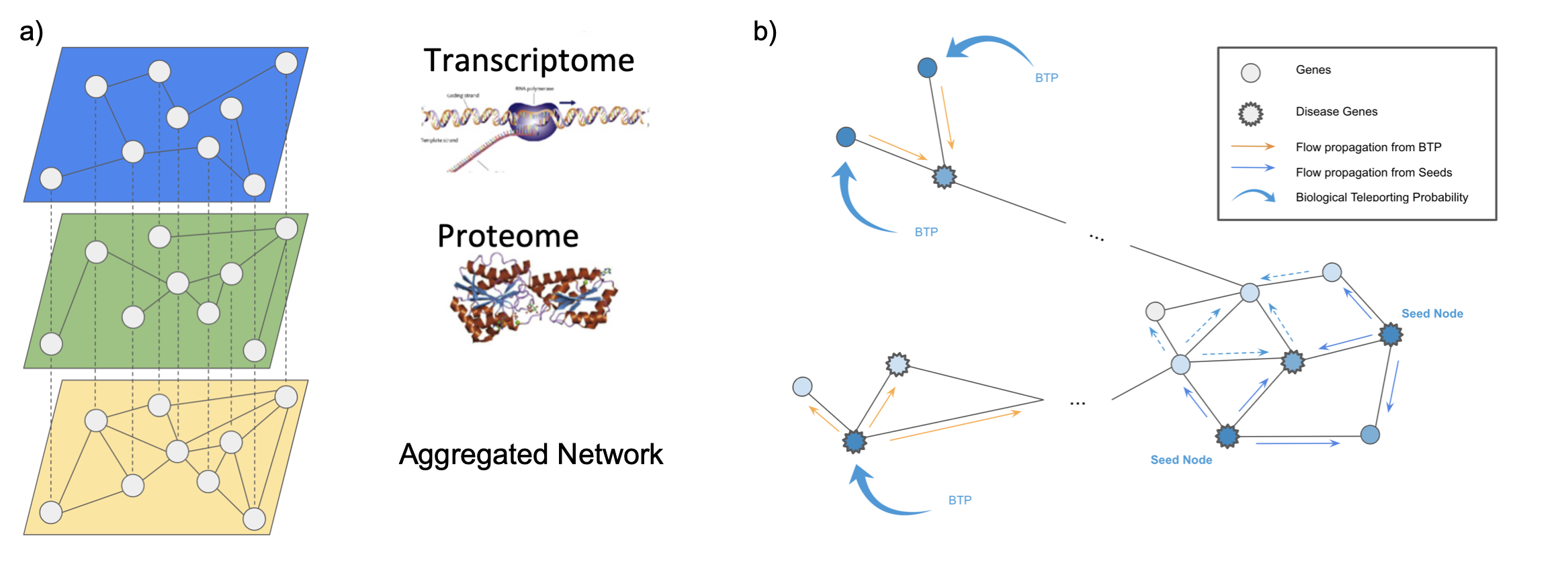}
	\caption{Network aggregation and network flow of Biological Random 
	Walks. a) shows the preprocessing step that combines PPI and 
	CO-Expression topology to derive a combined network. The transition 
	matrix of PPI and CO-Expression network are combined using a convex 
	combination to compute an aggregated transition matrix. Note that 
	what is termed ``aggregated network'' in tab a) is actually a 
	\emph{weighted} network that corresponds to the aggregated 
	transition matrix. b) Shows 
	the flow of the random walker when the personalization vector is 
	biased with disease specific information. In this way the flow can 
	propagate from seed nodes (i.e., known disease genes), but also 
	from node that are not part of the seed set but biologically 
	similar to it (i.e., biological teleporting probability (BTP)). }
    \label{fig:model_definition}
\end{figure*}

\subsection{Biological Random Walks}\label{se:bio_rw}
Biological Random Walks (BRW) build on the hypothesis that integrating different biological information sources may better reflect the complexity of protein interactions in a cell's process. In light of this insight, our approach integrates information on pairwise protein interaction of the Protein-Protein Interaction network (PPI) \cite{barabasi2011network} with other biological data in a unified framework. Our approach is to some extent agnostic to the particular biological data source, as long as it affords a principled notion of similarity between proteins. In the remainder, we use bold lowercase to denote vectors and capital, non-bold letters to denote matrices. Given a vector $\mathbf{x}$, $x_i$ denotes its $i$-th entry.

\paragraph{Notation.} We represent the PPI as an undirected graph $G= 
(V, E)$, with genes as vertices. Any edge $(i, j)$ represents a known 
protein-protein interaction recorded in the PPI. We assume $|V| = n$ in 
the following. For a given gene $i$, $N_h(i)$ denotes the subset of 
$G$'s vertices whose shortest path distance from $i$ is exactly $h$.

\subsubsection{Random Walks with Restart}
Random Walk with Restart (RWR) \cite{kohler2008walking} is a diffusion-based method, whose purpose is identifying disease modules that are topologically ``close'' to known disease genes in the interactome. It was shown to outperform other prioritization algorithms in many cases \cite{Navlakha}. In a nutshell, this algorithm can be seen as performing multiple random walks over the PPI network, each starting from a \textit{seed node} associated to a known disease gene, iteratively moving from one node to a random neighbour, thus simulating the diffusion of the disease phenotype across the interactome. More formally, the random walk with 
restart is defined as:

\begin{equation}
        \mathbf{p}^{(t+1)} = (1 - r) W \mathbf{p}^{(t)} + r \mathbf{q}.
\end{equation}

Here, $W$ is the column-normalized adjacency matrix of the graph andm$\mathbf{p}^{(t)}$ is a vector, whose $i$-th entry $p_i^{(t)}$ is the probability of the random walk being at node $i$ at the end of the $t$-th step. $r\in (0, 1)$ is the restart probability, i.e., the probability that the random walk is restarted from one of the (disease-associated) seed nodes in the next step. Upon a restart, the probability of restarting the random walk from some seed node $j$ is $q_j$. Vector $\mathbf{q}$ is normally called a \emph{personalization vector} in the Data Science literature. This random walk corresponds to an ergodic Markov chain \cite{levin2017markov} that admits a stationary distribution (i.e., a fixed point) $\mathbf{p}$. Nodes of the PPI are simply ranked by considering the corresponding entries of $\mathbf{p}$ in descending order of magnitude.

\subsubsection{Biological Information-Aware Random Walks}
%We next discuss the design of biased random walks that reflect heterogeneous biological information (Biological Random Walks).
For the sake of exposition, in the remainder we refer to the biological information associated to a gene (e.g., the set of its annotations) as the set of its \textit{features}. These can include (more precisely, be derived from) annotations from the Gene Ontology database \cite{DBLP:journals/nar/Consortium19a} (GO in the remainder) or gene expression levels. We remark that, in principle, any reliable information source on gene biology can be integrated. BRW ranks genes according to the main steps outlined below. 

Unlike \cite{kohler2008walking} and similar approaches, our method consists of two main steps: i) extracting statistically significant features from biological data, using them to compute a personalization vector and a transition matrix used by the BRW algorithm; ii) using the stationary distribution of the corresponding random walk to rank genes. Our approach to computing aggregated personalization vectors and transition matrices is outlined below, with further details given in Sections \ref{subse:teleport} and \ref{subse:trans_mat} of the supplementary material.

\paragraph{Computing a Personalization Vector.}Both gene annotation data and gene expression levels allow derivation of personalization vectors that reflect some notion of similarity between a gene and a disease, with the latter represented by the set of its seed genes. In the first case, this similarity is defined in terms of knowledge about genes' involvement in various biological functions and/or diseases. In the second case, similarity is defined in terms of co-expression levels of different genes in subject cases as opposed to expression levels in a control group, using data about a population of patients involved in a clinical trial.

\noindent
For annotation data, assume we have $\ell$ sources of biological information (e.g., GO, KEGG pathways etc.).Let $S$ denote the seed set. For every $j=1,\ldots, \ell$, we use $\source^j$ to denote the subset of annotations from the $j$-th source that are associated with at least one gene in $S$. Then, for every $j=1,\ldots, \ell$, we select a subset of annotations $\widehat{\source^j}$, filtering out  annotations that are not statistically significant, as shown in Supplementary Figure \ref{fig:enrichment_step} (i.e., $ p-value > 10^{-5}$, using Fisher Exact Test and FDR correction), so that $\source = \cup_{j=1}^{\ell}\widehat{\source^j}$ denotes the set of all statistically significant annotations that are associated with genes in $S$.Likewise, for every gene $i$ (not necessarily belonging to $S$), we denote by $\ann(i)$ the set of its annotations, possibly extracted from multiple biological information sources.  We assign each gene $i$ a weight $\weight_i$, which reflects the extent to which $i$ shares annotations that are statistically significant for genes that belong to the seed set of the disease under consideration. While other choices are possible, the definition we adopted reflects the extent of the \emph{inclusion} of $\ann_i$ in each of the $\ell$ sources:

\begin{equation}
	\weight_i =  
		\sum_{j=1}^{\ell} \frac{|\ann(i)\cap 
		\widehat{\source^j}|}{|\widehat{\source^j}|}.\footnote{It should be noted that \cite{kohler2008walking} corresponds to choosing
$\theta_i = 1$ if $i\in S$, $\theta_i = 0$ otherwise.}
\end{equation}

At this point, the components of a personalization vector $\pv$ can be computed as follows:

\begin{equation}
q_i = \frac{\theta_i}{\sum_{i = 0}^{N} \theta_i },
\end{equation}

with $N$ the number of genes we consider. Note that $q_i$ denotes the probability that, upon teleportation, the random walkers jumps to gene $i$. Further details are given in Section \ref{subsubse:pv} of the Supplementary material.

\noindent
We next discuss how to compute personalization vectors from gene 
expression data. An important goal is the identification of Differentially Expressed (DE) genes, whose expression levels systematically differ between case (Breast Cancer Cells) and a control group (Normal Breast Cells). We follow the approach proposed in \cite{menche2017integrating}, in which subjects of the control group are assigned personalized perturbation profiles (PEEPs), from which gene expression-aware personalization vectors can be derived.Succinctly put, for each gene $i$ and for each subject $j$, the expression level $l_i^j$ of gene $i$ in subject $j$ is compared with the distribution of the expression level of gene $i$ within the control group by taking the corresponding z-score $z_{ij}$. This approach allows association of a ``bar code'' to each subject. Following \cite{menche2017integrating}, we set $|z_{ij}| > 2.5$ as the threshold to declare gene $i$ differentially expressed in subject $j$. Following the general intuition stated in the introduction that disease genes generally are not scattered across the interactome, we also bias our choice towards differentially expressed genes that are closer to disease genes in the PPI. Eventually, we obtain a personalization vector $\pv$ that reflects both genes' differential expressions and vicinity to disease genes. 
For full details on computing PEEPs and deriving gene expression-aware personalization vectors, we refer the reader to Supplementary Section \ref{subsubse:pv}.

\paragraph{Computing a Transition Matrix.} Similar approaches can be used to derive a transition matrix for therandom walk with restart. Both approaches rely on the PPI, differing on the way PPI's edges are assigned weights that reflect genes' similarity and determine the probabilties of edge traversals. We leverage categorical, biological information (e.g., gene annotations)by defining a weighted transition matrix $\tmat$, in which each entry $W_{ij}$ depends on the extent to which nodes/genes $i$ and $j$ of the PPI share common annotations (i.e., they are involved in common biological processes) that are also significant for the disease. In more detail, considered genes $i$ and $j$, we define the following \emph{Disease Specific Interaction Function} (DSI function in the remainder):

\begin{equation}
    \dsi(i,j) = |\ann(i)\cap\ann(j)\cap\source|,
\end{equation}

where we remind that $\ann(k)$ denotes the set of gene $k$'s annotations, while $\source$ denotes the overall set of annotations that are statistically significant for disease genes.Intuitively, the higher $\dsi(i,j)$, the more $i$ and $j$ share annotations that are also statistically significant for the disease under consideration.In the end, edges in the PPI will be assigned weights depending on the DIS function as follows:

\begin{equation}
    \tmat_{ij} =
    \begin{cases}
     c + \dsi(i,j) & \text{if } (i,j) \in E, \\
      0        & otherwise.
    \end{cases}
  \end{equation}
  
Here, $c$ is a positive constant that accounts for usual sparsity of the available datasets, so that no biological information may be available for the end-points of a link in the PPI. In this case, the link receives a minimum weight $c$.As with personalization vectors, gene expression information about a population of patients can also be used to define a tissue-specific, population-dependent transition matrix. Specifically, gene expression information is used to assign weights to edges of the underlying PPI network, this time reflecting similarities between genes in terms of co-expression with respect to the subject population.Consider a CO-Expression network in which each pair of genes $(i,j)$ is assigned a score equal to the Pearson's correlation coefficient $pc_{i,j}$ between the expression levels of $i$ and $j$ within the population. We can define a transition matrix $\tmat$ by assigning each edge $(i, j)$ of the PPI network a probability as follows: 

\begin{equation}
    \tmat_{ij} = \frac{|pc_{ij}|}{\sum_{k \in \neigh(i)} |pc_{ik}|}
\end{equation}

where \textit{N(i)} is the set of \textit{i's} neighbors in the PPI network.Note that i) for every node/gene $i$ we consistently have a probability distribution over its incident edges and ii) the importance of edges reflects the absolute value of the correlation between the expression levels of two genes within the population of interest.

\paragraph{Integrating Biological Information and Gene Expression.} 
We discussed above two orthogonal approaches to the design of 
personalization vectors. The first one leverages similarities between 
the biological processes associated to known disease proteins and those 
to be prioritized. Hence, teleporting probabilities depend on the seed 
set through association with common biological processes.In 
the second case, teleporting probabilities depend on information that 
is tissue-specific (the level of expression in a population of subjects 
affected by a certain disease) and partly on the seed set, but this 
time through the PPI's topology. Hence, these two approaches largely 
rely on complementary sources of information. In order to integrate 
these complementary sources into a unique personalization vector that 
leverages both, we follow  a simple, yet mathematically principled 
approach, whereby we take a \emph{convex combination} of the 
corresponding personalization vectors. Namely, assume we have computed 
two personalization vectors $\pv_1$ and $\pv_2$, the former using 
biological information only, the latter using gene expression data and 
the PPI. We obtain a personalization vector as follows:
\begin{equation}
	\pv = \alpha\pv_1 + (1 - \alpha)\pv_2,
\end{equation}

where $\alpha\in [0, 1]$. Parameter $\alpha$ allows to weigh in the 
relative importance of the different information sources we are using. 
Intuitively, this type of aggregation amounts to considering a gene a 
potential candidate if it is statistically significant in terms of its 
involvement in biological processes, of its gene expression levels in 
the subject group, or both.\footnote{Note that this approach seamlessly 
extends to an arbitrary number of information sources.} The 
choice of $\alpha$ (and other parameters of the model), its impact on 
performance and dependence of the optimal choice on the scenario at 
hand are discussed in detail in Sections \ref{sec:experiments}, 
\ref{se:concl} and in Section \ref{subse:integration} of the 
supplementary material. \noindent\textbf{Remark.} Despite its 
simplicity, this is a mathematically principled choice. In particular, 
it is well-known \cite{jeh2003scaling} and easy to show that 
the stationary distribution corresponding to the convex combination of 
two personalization vectors $\pv_1$ and $\pv_2$ is itself the linear 
combination of the stationary distributions corresponding to $\pv_1$ 
and $\pv_2$ respectively. Briefly put, the parameter $\alpha$ allows us 
to tune the relative importance of the information providedby gene 
annotations and gene expression levels respectively. Moreover, this 
approach extends seamlessly to an arbitrary number of heterogeneous 
sources of information(and corresponding personalization vectors).
Other aggregation methods were also considered, yet they provided worse 
or at most comparable results. Some are presented in Section 
\ref{subsubse:int_pv} of the Supplementary material for the sake of 
completeness.  In the previous paragraphs, we have seen how we 
can derive random walk transition matrices using only information about 
biological processes (i.e., annotations) or gene expression data from a 
population of subjects. As with personalization vectors, multiple 
transition matrices derived from complementary biological sources can 
be integrated into a single \emph{aggregate transition matrix}. For 
example, assume we computed transition matrices $\tmat_1$ and $\tmat_2$ 
using biological and gene expression information respectively. Any 
convex combination of $\tmat_1$ and $\tmat_2$ is a feasible transition 
matrix. Namely, for some $\beta\in [0, 1]$, we consider the transition 
matrix
\[
	\tmat = \beta\tmat_1 + (1 - \beta)\tmat_2.
\]
It is easy to see that i) $\tmat$ is still a transition matrix, i.e., 
the entries of each row sum to one and that ii) the approach seamlessly 
extends to any number of complementary sources. In the case of 
biological annotations and gene expression data, $\beta = 1$ 
corresponds to only considering biological annotations, whereas $\beta 
= 0$ corresponds to only leveraging gene expression data. So, $\beta$ 
is a parameter, whose tuning allows us to weigh the importance of one 
source of information with respect to the other.

\section{Results}\label{sec:experiments}
This section investigates BRW's performance in prioritizing gene 
candidates for genetic diseases. 
%We analyzed the behavior of BRW  from 
%different perspectives. An in-depth, comparative analysis against other 
%algorithms on four well-studied cancer disease phenotypes and a corpus 
%of 70 mendelian diseases, comparing our framework against other 
%algorithms and understanding the sensitivity of BRW's performance to 
%different biological information sources.

\subsection{Experimental Setup}\label{subsection:data}
We used a number of biological data sources. Some of them were used as inputs, to define key parameters of our algorithms, while
others were used to biologically validate the results of the algorithms 
we considered. They are briefly described here and more extensively in 
Section \ref{se:apx_data} of the supplementary material.

\paragraph{Data sources.}
\noindent The experiments discussed in Section \ref{sec:experiments} were conducted on the HIPPIE Protein-Protein Interaction network (PPI) \cite{alanis2016hippie} and
on the same Protein-Protein Interaction network (PPI) 
as in \cite{ghiassian2015disease} for the sake of 
comparison. 

\noindent
We used three different sources of gene biological information: 
Gene Ontology Consortium\footnote{\url{http://geneontology.org/}.} 
where, for each gene, we downloaded its biological processes, KEGG \cite{doi:10.1002/pro.3715,10.1093/nar/gky962,10.1093/nar/28.1.27} and 
Reactome \cite{10.1093/nar/gkz1031}, which we used to download pathways' annotations. Gene expression datasets were downloaded from The Cancer Genome Atlas database\footnote{\url{https://portal.gdc.cancer.gov/}}. 

\noindent We validated the methods considered in this 
study both via an internal validation on known disease genes and 
through an external validation using drug target associations. In more 
detail, we used disease-gene associations as in 
\cite{ghiassian2015disease} that describe a corpus of 70 Mendelian 
diseases. From \cite{pinero2020disgenet}, we further derived known disease-gene associations for  
the four different cancer types that we investigate in Section 
\ref{sec:experiments} (i.e, breast cancer, lung adenocarcinoma, papillary thyroid cancer, colorectal adenocarcinoma).  Finally, we used 
Drug-Gene Target associations from 
DrugBank\footnote{\url{https://go.drugbank.com/}.} 
\cite{10.1093/nar/gkx1037}. We selected only drugs approved for Breast 
Cancer treatment from Food and Drug Administration 
\footnote{\url{https://www.cancer.gov/about-cancer/treatment/drugs/breast.}} 
(Supplementary Table \ref{tab:considered_drug}).

\noindent
All data sources mentioned above are more extensively discussed in 
Section \ref{se:apx_data} of the supplementary material.

\begin{figure*}[t]
    \centering
    \includegraphics[width=\textwidth]{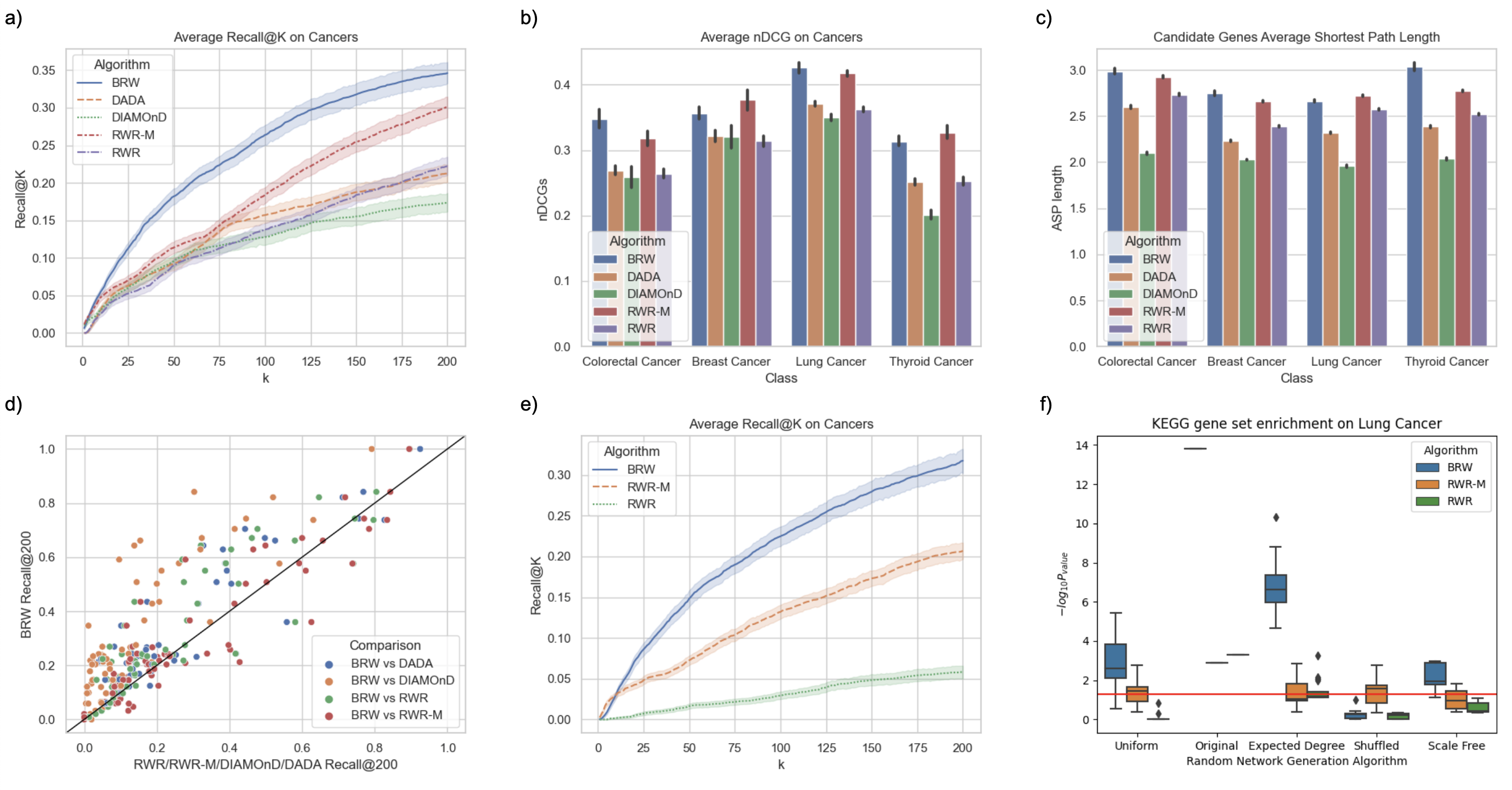}
    \caption{Algorithm comparison: a) and b) compare respectively the Recall@K and nDCG of some of well known network based approaches with BRW on four types of cancer using a 100-fold Monte Carlo random sampling validation choosing uniformly at random the 70\% of known disease genes and considering the remaining ones as the test set (30\%). c) shows the average shortest path length of top 200 candidates predicted by the analyzed algorithms. d) shows the comparison between BRW with the state of the art on a corpus of 70 Mendelian diseases downloaded from \cite{ghiassian2015disease}. e) compares multi-omics frameworks (BRW and RWR-M) on average Recall@k on cancer phenotypes when the PPI network is randomized. The average Recall@K is computed as described in the previous experiment. f) illustrate how multi-omics integration affect the bias induced by curated ontologies such as GO, Kegg, and Reactome and the PPI network. It compares the distribution of p-values (negative log scale) computed by the testsuit using KEGG pathways and \textbf{gseapy} on gene sets predicted on randomized PPI networks with those predicted on the original PPI. As hypothesized, BRW, having in input statistical significant pathways from KEGG, Reactome, and GO, returns biologically meaningful candidate sets. However, their significance is not comparable with the  set predicted using the original PPI.}
    \label{fig:computational_validation}
\end{figure*}

\paragraph{Baselines.} To provide a robust performance assessment, we compared BRW 
with a number of well-known, state-of-art baselines for disease 
gene prioritization, namely, RWR 
\cite{kohler2008walking}, DIAMOnD \cite{ghiassian2015disease}, DADA 
\cite{erten2011d} and RWR-M \cite{valdeolivas2019random}. For 
the sake of completeness, supplementary Section \ref{subse:new_vs_old} 
also compares our approach against an embryonic (and underperforming) 
version of our framework that was presented in \cite{gentili2019biological}.

\paragraph{Performance indices.} We used the widely adopted indices 
Recall@K and nDCG in some of the experiments described in the
remainder. To keep presentation self-contained, they are briefly 
described in Section \ref{apx:evaluation_scheme} of the supplementary 
material.

\subsection{Multi-Omics Integration Improves Algorithm 
performance}\label{subse:multi_omics}
In a first round of experiments, we performed two (internal) validation steps: i) we first 
compared the algorithms with respect to four cancer 
phenotypes (i.e, breast cancer, lung adenocarcinoma, papillary thyroid cancer, colorectal adenocarcinoma), for which both biological annotations and gene 
expression data are available; ii) as a further step, we performed a 
broader, yet less specific validation on a  
corpus of 70 manually curated mendelian diseases 
\cite{ghiassian2015disease}, for which only biological annotations were 
used. For each disease, we performed a mean 100-fold validation, by sampling 70\% of known 
disease genes uniformly at random  and using the rest to test the 
algorithms. For each tested disorder, we computed  
Recall@K and the nDCG of each algorithm.

\paragraph{Four cancer phenotypes.} We performed a grid search to identify the best 
combination of hyperparameters $(\alpha, \beta, r)$. The benchmark, 
discussed in Section \ref{subse:integration} of the supplementary 
material and illustrated in Supplementary Fig. \ref{fig:benchmarking_restart}  and 
Supplementary Fig. \ref{fig:benchmarking_alpha_beta}, highlights interesting, mixed 
results. For the task of disease gene prioritization\footnote{We remark 
that the settings discussed here are not optimal for other tasks in 
general, e.g., drug target discovery (see Section \ref{subse:breast}).} 
and to the purpose of computing personalization vectors (i.e., 
teleporting probabilities), the signal contained in statistically significant 
annotations derived from seed genes is definitely stronger
than the signal carried by gene expression, so that the best choice for 
disease gene prioritization  on cancer 
phenotypes is $\alpha = 1.0$, thus completely removing information 
about differentially expressed genes. However, gene expression 
information is crucial in determining the transition matrix of the 
random walk, with values of $\beta \in [0.25, 0.5]$ 
achieving best predictive performance, indicating that both 
information sources provide crucial information to BRW for disease gene 
ranking.

Results for the four cancer phenotypes show that algorithms that only leverage the PPI tend to perform worse in 
terms of both Recall@k and nDCG, as shown in figure 
\ref{fig:computational_validation} a) and \ref{fig:computational_validation} b). 
Conversely, multi-Omics methods that, like BRW and RWR-M, rely on 
multiple biological information sources typically perform better in 
terms of the aforementioned indices. Improvement of these methods over 
single-source baselines at least in part stems from the well-known fact that 
disease-associated genes tend to be involved in similar pathways and 
biological processes \cite{barabasi2011network}, see also Section \ref{subse:bias}.
%BRW and RWR-M combine multiple biological 
%sources in complementary ways. The former redistributes teleporting 
%probability mass across genes that tend to be involved in the same statistically 
%significant pathways as the seed genes; moreover, it aggregates the 
%topology of the interactome with the CO-Expression network. On the 
%other hand, RWR-M leverages a multiplex network that consists of a 
%Protein Interaction network, a CO-Expression network, and a Pathways 
%network built on top of different datasets (KEGG, Reactome). 
Interestingly, BRW's ability to bias the random walk towards related 
genes that do not necessary belong to the seed set seems to play an important 
role in improving prioritization of the test set, at least in terms of 
Recall@k. At the same time, 
RWR-M achieves similar performance (slightly better or worse, depending 
on the dataset) if one considers nDCG as a global measure of rank, as shown in figure 
\ref{fig:computational_validation} b). As previously remarked, biased teleporting 
makes BRW explore areas of the PPI network that could be relatively 
far from the seed set, a fact that is reflected in 
its candidate genes in the top 200 positions having higher 
average shortest path distance than other RWR-based  
methods that only teleport to genes of the seed set, as shown in Figure 
\ref{fig:computational_validation} c).

\paragraph{Mendelian diseases.} As a further internal validation, we provided a less specific yet 
broader, comparative assessment of BRW, by performing a Monte Carlo 
cross-validation \cite{dubitzky2007fundamentals} on ``gold standard'' 
gene sets. These sets contain known genes associated with 70 diseases, 
which were previously selected in \cite{ghiassian2015disease} from 
OMIM and PheGenI databases. As discussed more in detail at the end of 
Section \ref{se:apx_ppi}, in this experiment (and only in this one) we 
used the PPI considered in \cite{ghiassian2015disease}, for the sake 
of consistency.\footnote{All other experiments use the more recent 
HIPPIE-v2.2 PPI network of \cite{alanis2016hippie}.} In this second round of experiments, 
we set $\alpha = 1.0$ and $\beta = 1.0$, since gene expression is not used. 
Figure \ref{fig:computational_validation} 
d) compares the performances of RwR, DIAMOnD, DADA, RWR-M and BRW in 
terms of Recall@k. 
Our heuristic ranks more known disease 
proteins than DIAMOnD in the top 200 positions for 95 percent of 
disorders analyzed and respectively for 72, 73, and 63 
percent of the disorders for RwR, DaDA, and RWR-M.

\subsection{Randomization and Bias}\label{subse:bias}

 \begin{figure*}[t]
    \centering
    \includegraphics[width = \textwidth]{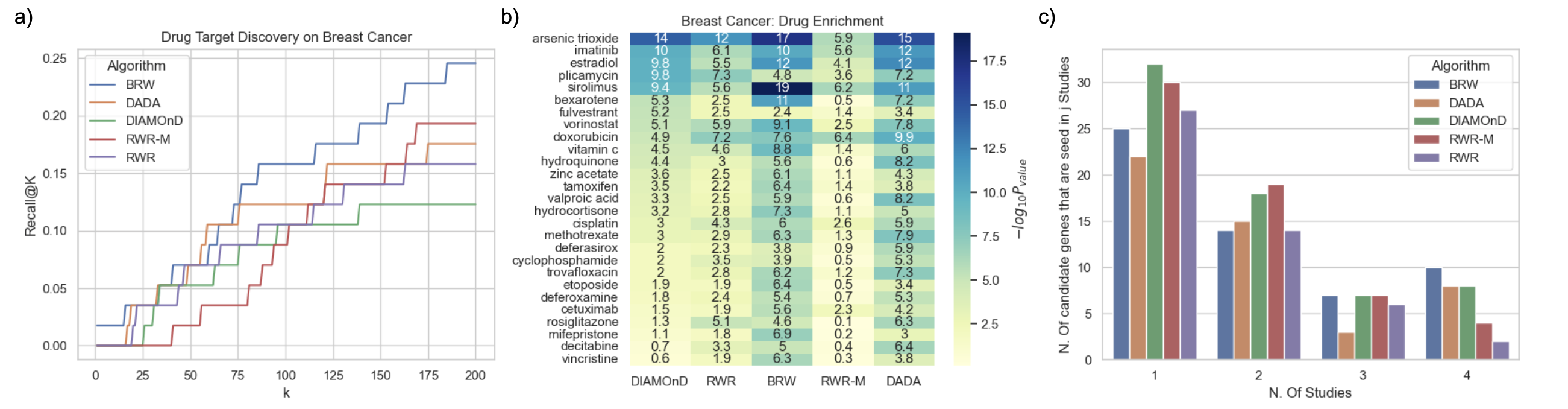}
    \caption{\textbf{Algorithm Comparison on Breast Cancer Disease:} a) 
    percentage of Breast Cancer Drug targets found by each framework in 
    the top K positions. b) drugs that are enriched (corrected p-value 
    < $10^{-5}$) in at least one of the predicted candidate gene sets. 
    c) number of candidate genes, predicted by each algorithm, that are 
    frequently mutated in at least j of the remaining studies, with $j \in \{1, 2, 3, 4\}$.}
    \label{fig:breast_cancer_validation}
\end{figure*}

Biological Random Walks and RWR-M leverage multiple data sources. As a result, we 
expect their results to be less affected by random noise in the PPI 
with respect to other heuristics. To test this 
hypothesis, we performed a first experiment, by performing an internal 
validation as done in Section \ref{subse:multi_omics}, but this time running the random 
walk-based heuristics using degree-preserving randomized version of the 
PPI. To this purpose, we 
implemented the degree-preserving randomization algorithm of
\cite{milo2003uniform}, also described in Section 
\ref{subse:apx_random} of the supplementary material. As 
hypothesized (Fig. \ref{fig:breast_cancer_validation} e)), by leveraging 
multiple biological sources, BRW and RWR-M are less affected by randomization of the PPI. 
This effect is stronger in BRW, whose teleporting probabilities also depend 
on phenotype information (statistically significant ontologies, 
pathways, and differentially expressed genes). This is further shown in the more 
detailed Supplementary Fig. \ref{fig:robustness_and_bias}  a) of the supplementary 
material, highlighting a positive correlation between the number of test genes ranked in the 
first 200 positions (Recall@K) and the value of BRW's restart 
probability $r$. 
%When $r$ increases, the random walker will be 
%teleported more frequently to nodes with high biological similarity to 
%the seed set and rely less on the aggregated network topology (i.e., 
%bypassing the randomized PPI network). 
As remarked above, this effect is also present in RWR-M, in which  
heterogeneous biological sources are summarized in different, layered 
networks. In particular, as shown in Supplementary Fig. \ref{fig:robustness_and_bias} 
a) of the supplementary material, RWR-M's performance degrades 
significantly if one also randomizes the Pathways network used by the 
algorithm. Altogether, these results suggest that phenotypical information associated to 
the nodes plays an important role in prioritization for these 
heuristics, owing to the fact that known disease 
proteins tend to be involved in the same pathways, a fact that becomes 
apparent in an internal-only validations as the one considered 
in this section.

%\review{This experiment pointed out a critical feature of integrating 
%several data sources: the combination between PPI network and 
%ontologies can moderate the noise that might affect one or more 
%biological information. However, the bias induced by up-to-date PPI 
%networks and manually curated ontologies such as GO, KEGG, and Reactome 
%might produce more biological meaningful candidate genes when an 
%algorithm runs on a random PPI network than on the original one. 
%Indeed, \cite{lazareva2021limits} has shown that several Active Module 
%Identification Methods inherit the bias induced by the PPI network and 
%predict more biologically meaningful candidate gene sets when they run 
%on preserving-degree randomized PPI networks.} 

The results above are part of a more general phenomenon. As 
shown in \cite{lazareva2021limits}, several heuristics inherit biases 
present in up-to-date PPI networks and manually curated ontologies such 
as GO, KEGG, and Reactome, in some cases with results that improve when 
the PPI is replaced by a randomized one. To further investigate this 
issue, in particular, the bias inherited by BRW when ontologies from 
manually curated data sources are used, we considered the AMIM (Active Module 
Identification Methods) test suite proposed in 
\cite{lazareva2021limits}, which allows systematic comparison of 
candidate gene sets predicted using an original PPI network and 
perturbed versions thereof. To this purpose, we considered the Lung 
Cancer phenotype, which was both considered in Section 
\ref{subse:multi_omics} and is used as a benchmark in 
\cite{lazareva2021limits}. We 
considered randomized versions of the original PPI obtained using all 
randomization algorithms implemented in the test suite (i.e., expected 
degree, uniform, shuffled and scale-free). 
Candidate genes ranked by BRW, RWR-M, and RWR using the original PPI as 
input were compared with the results obtained using each of the 
aforementioned randomized counterparts of the PPI. Statistical 
significance was computed on KEGG pathways enrichments computed by 
the test suite on predefined KEGG pathway.

Fig. \ref{fig:computational_validation} f) shows that BRW, 
RWR-M, and RWR extract statistically significant candidate gene sets on 
the original PPI network. Relying only on the PPI, RWR provides no 
statistically significant results when the PPI is randomized, except 
when expected degree is preserved.\footnote{\label{foot:stat}This phenomenon arises 
because the stationary distribution of a random walk on an 
undirected network follows exactly the degree distribution. 
As a result, the stationary distribution of a random walk with restart over an undirected network is 
positively correlated with degree distribution.} On the other hand, 
BRW and RWR-M inherit PPI and ontology's biases, so that their results 
are still statistically significant when the PPI is randomized using 
expected degree, scale-free and uniform randomization algorithms. Still, BRW's results are 
considerably more significant when the original network is used, 
indicating that topology of the PPI is crucial to (more) effectively 
propagate information extracted from other biological sources. BRW 
predicts more significant outcomes on the actual PPI than on the randomized versions. 
While all ontologies (GO, KEGG, and Reactome) were considered in the 
experiment summarized in Fig. \ref{fig:computational_validation} f), we 
further investigated BRW's behaviour when KEGG (which is used in 
the enrichment) is not used. In this case, BRW's results are no longer 
statistically significant if one randomizes the network, with the 
exception of a mild significance when 
expected degree is preserved (see again footnote \ref{foot:stat} for an 
explanation).

\subsection{A Case Study: Breast Cancer Phenotype}\label{subse:breast}
In this section, we discuss the results of an in-depth analysis of BRW's 
performance on the Breast Cancer phenotype, a global health concern
\cite{siegal2014cancer,crimini2021precision}, with 284,200 new 
cases and more than 44,000 deaths in the USA in 2021 
\cite{siegal2014cancer}. In particular, we present 
i) an external validation using drugs FDA approved 
drugs; ii) Drug enrichment; iii) an assessment of the algorithm(s) stability across multiple 
populations.

%Breast cancer (BC) represents a global health concern, with 284,200 new 
%cases and more than 44,000 deaths in the USA in 2021 
%\cite{siegal2014cancer}. Extensive characterization of genetic and 
%epigenetic abnormalities in breast cancer has led to novel biomarkers 
%and targeted therapies. Unfortunately, patient-drug matching still 
%fails in a significant number of cases \cite{crimini2021precision}. In 
%this scenario, network-based approaches can provide useful insight into 
%key biological mechanisms and guide the quest for novel, more specific 
%therapeutic targets. This section presents an in-depth analysis of BRW 
%and its performance on the Breast Cancer phenotype using  multiple 
%biological information sources, along with a comparison with other 
%disease gene prioritization algorithms.

\subsubsection{Drug Target Discovery}\label{subsub:target}

We validated the top candidate gene sets prioritized by each algorithm 
along different axes. To this purpose, we created a test set of target 
genes for Breast Cancer drugs approved by the FDA as described in 
Section \ref{subsection:data}. Results are summarized in Figure 
\ref{fig:breast_cancer_validation} a), showing Recall@K for the algorithms 
we considered. Compared to other baselines, BRW prioritizes the highest number (25\%) of drug 
targets in the top 200 candidates. 
%The higher the alpha value, the 
%higher the percentage of genes predicted, thus suggesting that drug 
%targets are more correlated with statistically significant pathways 
%than gene expression data. Furthermore, we can notice that the best 
%performance is obtained with a low restart probability. 
Looking 
at the specific gene targets predicted by the algorithms, BRW predicts 
the highest number of genes (14 out of the 20 predicted genes), making 
four unique predictions, namely, Cyclin Dependent Kinases 4 and 6 
(\textit{CDK4}, \textit{CDK6}), the Protein Kinase C Zeta 
(\textit{PRKCZ}) and Caspase 3 (\textit{CASP3}). On the other hand, the DNA 
Topoisomerase II Alpha (\textit{TOP2A}) and the progesterone receptor 
(\textit{PGR}) genes are uniquely predicted by DIAMOnD. The Protein 
Kinase C Theta gene (\textit{PRKCQ}) is only predicted by RWR-M 
algorithm. More details are given in Supplementary Table 5.

Furthermore, we validated the prioritized genes from a 
different perspective: we considered the group of drug targets returned 
by each framework and showed the drugs that target them. We filtered 
out drugs in Drug Bank that were not annotated with the "Breast Cancer" 
or related associated condition. Supplementary Table 
\ref{tab:drug_enrichment_benchmark_best_combination} shows how the 
number of drugs ranges from 6 to 17 as we change the combination of 
hyper-parameters. As expected, the number of 
drugs correlates positively with the predicted drug target percentage 
(Recall@200). Supplementary Table \ref{tab:drug_discovered_table} shows 
the drugs prioritized (i.e., a drug with at least one drug target 
prioritized in the top 200 positions) by BRW and the other algorithms. 
BRW prioritizes the highest number of drugs. Interestingly, CDK4/6 inhibitors 
palbociclib, ribociclib, and abemaciclib, currently used to treat 
hormone receptor-positive/HER2-negative metastatic breast cancer 
\cite{duranti2021breast}, are only predicted by BRW.

Finally, we identified drugs that have an enrichment with the top 
candidates genes predicted by each algorithm. To identify enriched drugs, we 
used the \textbf{gseapy} package for gene set enrichment analysis 
\cite{Subramanian15545}, and we chose drugs that were enriched by at 
least one algorithm with a corrected p-value lower than $10^{-5}$. 
In particular, Supplementary Table 
\ref{tab:drug_enrichment_benchmark_best_combination} shows how the 
enrichment is affected by various hyper-parameters combinations. 
In this case, drug enrichment correlates positively with gene 
expression, and the best combination is obtained when $\alpha$ and 
$\beta$ are equal to $0.25$. Indeed, while drugs target 
and inhibit genes involved in disease-specific pathways, the effect of 
the drugs can be measured by the differential expression and the 
co-expression between targets and nearby genes \cite{chen2017reversal}. 
In general, results highlight the following trends: i) different biological sources 
(in our case, gene expression and ontologies) provide complementary 
information, with different subsets of FDA approved drug targets 
significantly enriched for both low (gene expression bias) and high 
(ontologies bias) values of the parameters $\alpha$ and $\beta$; ii) 
best performance is achieved for lower values of the restart 
probability $r$ ($0.25$), confirming that BRW's prioritization depends 
on information that is not necessarily confined to the immediate 
neighbourhood of the seed set.
\medskip

Fig. \ref{fig:breast_cancer_validation} b) highlights drugs that are 
enriched in BRW and in the other baselines considered in this 
manuscript, with details for each drug reported in Supplementary Table 6. Interestingly, 11 FDA-approved drugs for breast cancer treatment are enriched in BRW gene candidates (i.e., fulvestrant, doxorubicin, paclitaxel, tamoxifen, methotrexate, letrozole, cyclophosphamide, trastuzumab, fluorouracil, gemcitabine).  The mTOR inhibitor sirolimus showed the best significance. While it is not currently used to treat breast cancer, preclinical \textit{in vivo} studies  demonstrated its potent antiangiogenic activity on breast cancer models \cite{muhammad2022rapamycin}

\subsubsection{Breast Cancer - Multi Population Study}\label{subse:pop_studies}
A desirable property of disease gene prioritization should be a certain stability in the set of proposed disease gene candidates across different populations. In other words, to some extent (e.g., up to intrinsic biases or qualitative differences in the datasets used), results should not overly depend on the specific dataset the algorithm is analyzing. To investigate this aspect, we characterized the behaviors of BRW and the other baselines we considered when applied to different population studies on Breast Cancer.
 \noindent We considered Invasive Breast Cancer population studies retrieved from cBioPortal datasets  \cite{stephens2012oslo,shah2012clonal,banerji2012sequence,kan2018multi,ciriello2015comprehensive} (see Section \ref{se:apx_data} of supplementary material), which allowed us to identify five different seed sets for the algorithms we considered, one per study. Figure \ref{fig:breast_cancer_validation} d) plots the number of candidate genes, predicted by each algorithm, that are seeds in at least \textit{j} other studies (frequently mutated genes in the associated populations, i.e. mutation frequency > 1\%), for $j \in \{1,2,3,4 \}$. Notably, BRW has the highest number of candidates that are frequently mutated in the other 4 studies (the far right blue bar). The 10 genes identified by BRW, that are frequently mutated in 4 out of the 5 breast cancer populations are \textit{AR}, \textit{JUN}, \textit{STAT3}, \textit{NOTHC1}, \textit{JAK2, HDAC1, SMAD4, YAP1,} and \textit{CHD4}. While the Histone Deacetylase 1 (\textit{HDAC1}) is also retrieved by all the other algorithms, \textit{SMAD4} and \textit{CHD4} are only predicted by BRW. The remaining genes are returned by BRW and at least another heuristic. Details are reported in Supplementary Table 7.

\section{Discussion}\label{se:concl}
Guided by the hypothesis that disease causing genes often share important common 
biological processes and pathways, we extended the random 
walk with restart approach to disease gene prioritization, proposing a 
framework that allows seamless integration of multiple biological 
information sources. The proposed approach consists of two main steps: i) extracting 
significant disease features from disease-term association data, 
such as statistically significant biological processes and 
pathways, and ii) using these features to bias the random walk with 
restart in a way that is consistent with the biological sources used. 
These two aspects are discussed in Section 
\ref{section:materials_methods}.

In general, BRW outperforms, in terms of standard indices of predictive 
accuracy such as recall and nDCG, previous 
frameworks that only rely on a s single biological source, such as Random 
Walks with Restart \cite{Navlakha}, DaDa \cite{erten2011d} and 
DIAMOnD \cite{ghiassian2015disease} that rely on Protein-Protein 
Interaction network. This is also true for an extension of RWR, namely, RWR-M 
\cite{valdeolivas2019random}, that performs a random walk on a multi-layer 
network. 
Using a Monte Carlo random sampling validation, we 
showed that prioritization results returned by BRW frequently 
outperform other baselines on four different cancer types: 
Breast, Colorectal, Lung, and Thyroid cancer. These results were 
further supported by a broader, 
computational validation on a corpus of seventy Mendelian disease 
manually curated by \cite{ghiassian2015disease}.

A main aim of precision medicine is to use disease genes to enable tailored 
treatments. In this perspective, we investigated how the candidate genes prioritized by each 
framework are related to Breast Cancer drug targets. Results show that 
BRW prioritizes the highest number of drug targets in the top 200 candidates 
(Figure 3a). As a further assessment, we considered the 
group of drug targets returned by each algorithm and identified the drugs 
that target them. We found that BRW prioritizes the highest 
number of FDA-approved drugs for Breast Cancer treatment (Supplementary 
Table 4). 
In general, we noticed that drug targets are more correlated with phenotypic 
pathways, while drug enrichment highlights how drugs affect gene 
expression in terms of co-expression and differential expression. This 
is not surprising: indeed, although drugs often target and inhibit 
genes involved in disease-specific pathways, the effect of the drugs 
can be measured by the differential expression and the co-expression 
between targets and nearby genes \cite{chen2017reversal}. 

To investigate the stability of the proposed disease gene candidates, 
we characterized the behaviors of BRW and the previous frameworks 
across different population studies. We selected “invasive breast 
cancer” as a phenotype and we retrieved data for 5 different populations from 
cBioPortal (https://www.cbioportal.org/). Consistently, BRW showed 
excellent stability, with the highest number of gene candidates in one 
study that are frequently mutated in the other four (Figure 
\ref{fig:breast_cancer_validation} c), Supplementary Table 7). 

The BRW framework is not without 
limitations. On one hand, inducing a bias in teleporting probability and the 
transition matrix through the use of ontologies improves predictive 
accuracy of the algorithm. On the other hand, this bias can hinder 
BRW's ability to identify new disease-related pathways. For this 
reason, we believe it is important to exploit the framework's ability 
to integrate heterogeneous, hopefully complementary, data sources. Furthermore, experiments summarized in Fig. 
\ref{fig:benchmarking_alpha_beta} and Supplementary Table 
\ref{tab:drug_enrichment_benchmark_best_combination} quantitatively 
showed that every data source comes with its bias. As a result, the relative 
weights attributed to different biological information sources 
(the combined choice of $\alpha$ and $\beta$ in our case) can 
significantly affect predictive accuracy, with a magnitude that in 
general depends on the validation test used. For example, while 
ontologies showed most effective in a computational validation and for 
drug target discovery, gene expression proved particularly useful in 
identifying a candidate gene set that is highly enriched in breast 
cancer and cancer related drugs. In our opinion, these results provide support to the 
use of (at least partially) complementary sources of biological 
information, even though these sometimes present non-negligible 
correlations, as discussed in Section \ref{subse:bias}.

%A couple of takeaways are the following: i) the 
%relative weights attributed to different biological information sources 
%(the combined choice of $\alpha$ and $\beta$ in our case) can have 
%a non-negligible impact on prediction accuracy; ii) interestingly, the 
%best combination of these parameters depends the validation test, indicating that the predictive power of 
%different biological information sources varies across different 
%scenarios. In our opinion, these results provide support to the 
%use of (at least partially) complementary sources of biological 
%information, even though these sometimes present non-negligible 
%correlations, as discussed in Section \ref{subse:bias}.

We also emphasize that, in this study, we only leveraged a limited set 
of disease-specific data sources (co-expression and differential 
expression). Potential performance improvements might be achieved by 
integrating further disease information sources, such as methylation data, microRNA expression, or microRNA-target 
gene associations. Doing this might provide further insights into key 
biological mechanisms and provide new prospective gene targets, though, 
of course, only functional studies can provide the ultimate answer as 
to their biological role.

\section{Funding.}
Partially supported by the ERC Advanced Grant 788893 AMDROMA 
``Algorithmic and Mechanism Design Research in Online Markets'', the EC 
H2020RIA project ``SoBigData++'' (871042), and the MIUR PRIN project 
ALGADIMAR ``Algorithms, Games, and Digital Markets''.
\bibliographystyle{alpha}
\bibliography{references}

\appendix

\newpage
\setcounter{figure}{0}
\renewcommand{\figurename}{Supplementary Fig.}

\begin{appendices}
\section{Data}\label{se:apx_data}

\subsection{Protein-Protein Interaction networks}\label{se:apx_ppi}
To run the experiment presented in Section \ref{sec:experiments}, we considered two different PPI networks:
\begin{enumerate}
\item HIPPIE-v2.2: Protein-protein interaction network downloaded from 
\cite{alanis2016hippie}. We filtered the network, keeping only the 
edges with a score larger than 0.7. Then, we mapped each node to its 
Ensembl id, removing nodes mapped on multiple Ensembl Ids. This way, we derive a PPI network with 88,865 edges and 12,148 nodes.

\item \cite{ghiassian2015disease}: We used the same Protein-Protein 
Interaction network (PPI) as in \cite{ghiassian2015disease}. In 
\cite{ghiassian2015disease}, the authors only considered direct 
physical protein interactions with reported experimental evidence. They 
derive this PPI network with the help of several data sources: 
TRANSFAC, \cite{kel2003transfac}, IntAct \cite{Armean2009IntAct}, 
MINT \cite{Aryamontri2009mint}, BioGRID \cite{aryamontri2010Biogrid}, HPRD \cite{Venugopal2008HPRD} KEGG, BIGG \cite{Schellenberger2010BIGG}, and CORUM \cite{Ruepp2009CORUM}. Then, we considered the main connected component of 
the network and we removed self-loops (i.e., edges describing proteins'
self-interactions). The resulting graph consists of 13396 nodes and 
138405 edges. 

\end{enumerate}
\paragraph{Remark.} Note that the PPI of \cite{ghiassian2015disease} 
was used only in the internal validation on the 70 mendelian diseases, 
whose results are discussed in Section 3.2 of the main article. This 
was done for consistency, since the network considered in 
\cite{ghiassian2015disease} was explicitly curated to include all 
genes that are associated to at least one of the 70 diseases under 
consideration, some of which might be missing in other, even more 
recent, networks. Instead, all other experiments in this paper used the 
more recent HIPPIE-v2.2 PPI \cite{alanis2016hippie}.

\subsection{Biological data sources}
\paragraph{Gene annotations.}
We used three different sources of gene biological information: 
Gene Ontology Consortium\footnote{\url{http://geneontology.org/}.} 
where, for each gene, we downloaded its biological processes, KEGG 
Pathways \cite{doi:10.1002/pro.3715,10.1093/nar/gky962,10.1093/nar/28.1.27} and 
Reactome \cite{10.1093/nar/gkz1031}, which we used to download 
pathways' annotations.

\paragraph{Pathways Network:}
a network connecting proteins according to pathway interaction data, using the R package
graphite \cite{sales2012g}. 
\paragraph{Tissue Network:}a co-expression network from RNA-Seq data downloaded from the Human Protein Atlas (http://www.proteinatlas.org)
(Uhlen et al., 2015). We computed Spearman correlations of TPM (transcript per million) expression
data from 32 tissues and 45 cell lines, and selected correlation having an absolute value $\ge$ 0.7 to be included in the network.

\paragraph{Gene expression data (TCGA).}
Gene expression datasets were downloaded from
The Cancer Genome Atlas datasets.\footnote{\url{https://portal.gdc.cancer.gov/}}. We collected from CDC portal the RNA-Seq data of several type of cancers. Table \ref{tab:TCGA} describe the cancer types downloaded and the number of Primary Tumor and Solid tissue normal used to compute the set of differentially expressed genes,and co-expression network.

\begin{table}[htp]\centering
\caption{The table shows all the TCGA projects downloaded to create the set of differentially expressed genes and the CO-expression network}\label{tab:TCGA}
\scriptsize
\begin{tabular}{l|r|r|r|r}
\textbf{Project ID} &\textbf{Cancer Type} &
\textbf{N. of Primary Tumor} &\textbf{N. of Solid Tissue Normal} \\
\textbf{TCGA-BRCA} &Breast Cancer &1077 &99 \\
\textbf{TCGA-LUAD} &Lung Adenocarcinoma &519 &58 \\
\textbf{TCGA-THCA} &Papillary Thyroid Cancer &497 &56 \\
\textbf{TCGA-COAD} &Colorectal Adenocarcinoma &456 &41 \\
\end{tabular}
\end{table}

\subsection{Data sources for experimental validation}

\paragraph{Disease-gene associations.}We used disease-gene associations from 
\cite{ghiassian2015disease}. Out of a corpus of 
gene-disease associations involving 69 well-characterized diseases retrieved from OMIM\footnote{\url{https://www.ncbi.nlm.nih.gov/omim/}} (Online Mendelian 
Inheritance in Man;  \cite{OMIM}) and 
PheGenI\footnote{\url{https://www.ncbi.nlm.nih.gov/gap/phegeni}}  (Phenotype-Genotype Integrator; \cite{10.1038/ejhg.2013.96}) databases, we focused on the 
\textbf{Breast Cancer} phenotype, which has associations with 40 genes.

\paragraph{Disease-Gene associations of Invasive Breast Cancer on 
several population studies.} From 
\textbf{cBioPortal}\footnote{\url{https://www.cbioportal.org/}} we downloaded Invasive Breast Cancer mutated genes belonging to 5 studies \cite{stephens2012oslo,shah2012clonal,banerji2012sequence,kan2018multi,ciriello2015comprehensive} (Some information about the subjects sequenced by each study is reported in Supplementary Fig. \ref{fig:population_study_table}) and we filtered out genes mutated in less than 1\% of the studied population and that are not present in the OncoKB \cite{chakravarty2017oncokb}. More precisely, for each study (\emph{reference population}), a reliable seed set was obtained, by keeping genes that i) were highly mutated in the population study under consideration (mutation frequency $> 1$\%) and ii) were known cancer genes, present in OncoKB database  \cite{chakravarty2017oncokb}. At the same time, significantly mutated genes identified across the remaining four studies provided a natural validation set. Thus, for each algorithm and for each reference population, a candidate gene prioritized by the algorithm is considered to be a potential disease gene if it appears mutated gene in at least one of the other population studies.
The underlying hypothesis is that genes involved in a pathological pathway are likely to be mutated in at least one population.

\begin{table}[!htp]\centering
\caption{Breast cancer drugs filtered from DrugBank to derive a set of target genes we used to validate algorithm prediction on an external dataset. }\label{tab:considered_drug}
\scriptsize
\begin{tabular}{lp{2.5cm}p{2.5cm}|p{2.5cm}p{2.5cm}p{2.5cm}}
\textbf{Drug Bank ID} &\textbf{Drug Name} &\textbf{Target Genes} &\textbf{Drug Bank ID} &\textbf{Drug Name} &
\textbf{Target Genes} \\
\textbf{DB12001} &Abemaciclib &CDK4, CDK6 &\textbf{DB05773} &Trastuzumab emtansine &ERBB2 \\
\textbf{DB01229} &Paclitaxel &TUBB1, BCL2, MAP4, MAP2, MAPT, NR1I2 &\textbf{DB09037} &Pembrolizumab &PDCD1 \\
\textbf{DB05773} &Trastuzumab emtansine &ERBB2 &\textbf{DB11730} &Ribociclib &CDK4, CDK6 \\
\textbf{DB01590} &Everolimus &MTOR &\textbf{DB01259} &Lapatinib &EGFR, ERBB2 \\
\textbf{DB12015} &Alpelisib &PIK3CA &\textbf{DB01006} &Letrozole &CYP19A1 \\
\textbf{DB01217} &Anastrozole &CYP19A1 &\textbf{DB09074} &Olaparib &PARP1, PARP2, PARP3 \\
\textbf{DB00282} &Pamidronic acid (Aredia) &FDPS, GGPS1, CASP3, CASP9 &\textbf{DB00351} &Megestrol acetate &PGR, NR3C1 \\
\textbf{DB01101} &Capecitabine &TYMS &\textbf{DB00563} &Methotrexate &DHFR, TYMS, ATIC \\
\textbf{DB00531} &Cyclophosphamide &NR1I2 &\textbf{DB11828} &Neratinib &EGFR \\
\textbf{DB01248} &Docetaxel &TUBB1, BCL2, MAP4, MAP2, MAPT, NR1I2 &\textbf{DB09074} &Olaparib &PARP1, PARP2, PARP3 \\
\textbf{DB00997} &Doxorubicin &TOP2A, NOLC1 &\textbf{DB01229} &Paclitaxel &TUBB1, BCL2, MAP4, MAP2, MAPT, NR1I2 \\
\textbf{DB00445} &Epirubicin &TOP2A &\textbf{DB09073} &Palbociclib &CDK4, CDK6 \\
\textbf{DB14962} &Trastuzumab deruxtecan &FCGR1A, TOP1 &\textbf{DB00282} &Pamidronic acid &FDPS, GGPS1, CASP3, CASP9 \\
\textbf{DB00445} &Epirubicin &TOP2A &\textbf{DB09037} &Pembrolizumab &PDCD1 \\
\textbf{DB08871} &Eribulin &BCL2, TUBB1 &\textbf{DB06366} &Pertuzumab &ERBB2 \\
\textbf{DB01590} &Everolimus &MTOR &\textbf{DB14740} &Hyaluronidase &TGFB1 \\
\textbf{DB00990} &Exemestane &CYP19A1 &\textbf{DB12015} &Alpelisib &PIK3CA \\
\textbf{DB00544} &Fluorouracil &TYMS &\textbf{DB12893} &Sacituzumab govitecan &TACSTD2, TOP1, FUBP1 \\
\textbf{DB14962} &Trastuzumab deruxtecan &FCGR1A, TOP1 &\textbf{DB00675} &Tamoxifen &ESR1, ESR2, EBP, AR, KCNH2, NR1I2, ESRRG, SHBG, MAPK8, PRKCA, PRKCB, PRKCD, PRKCE, PRKCG, PRKCI, PRKCQ, PRKCZ \\
\textbf{DB00539} &Toremifene &ESR1, SHBG &\textbf{DB11760} &Talazoparib &PARP1, PARP2 \\
\textbf{DB00947} &Fulvestrant &ESR1 &\textbf{DB11595} &Atezolizumab &CD274 \\
\textbf{DB01006} &Letrozole &CYP19A1 &\textbf{DB00539} &Toremifene &ESR1, SHBG \\
\textbf{DB00544} &Fluorouracil &TYMS &\textbf{DB00563} &Methotrexate &DHFR, ATIC, TYMS \\
\textbf{DB00441} &Gemcitabine &RRM1, TYMS, CMPK1 &\textbf{DB12893} &Sacituzumab govitecan &TACSTD2, TOP1, FUBP1 \\
\textbf{DB00014} &Goserelin &LHCGR, GNRHR &\textbf{DB11652} &Tucatinib &ERBB2, ERBB3 \\
\textbf{DB08871} &Eribulin &BCL2, TUBB1 &\textbf{DB01259} &Lapatinib &EGFR, ERBB2 \\
\textbf{DB00072} &Trastuzumab &ERBB2 &\textbf{DB12001} &Abemaciclib &CDK4, CDK6 \\
\textbf{DB09073} &Palbociclib &CDK4, CDK6 &\textbf{DB00570} &Vinblastine &TUBA1A, TUBB, TUBD1, TUBG1, TUBE1, JUN \\
\textbf{DB00441} &Gemcitabine &RRM1, TYMS, CMPK1 &\textbf{DB01101} &Capecitabine &TYMS \\
\textbf{DB04845} &Ixabepilone &TUBB3 &\textbf{DB00014} &Goserelin &LHCGR, GNRHR \\
\end{tabular}
\end{table}

\begin{figure}
\centering
\includegraphics[width = \textwidth]{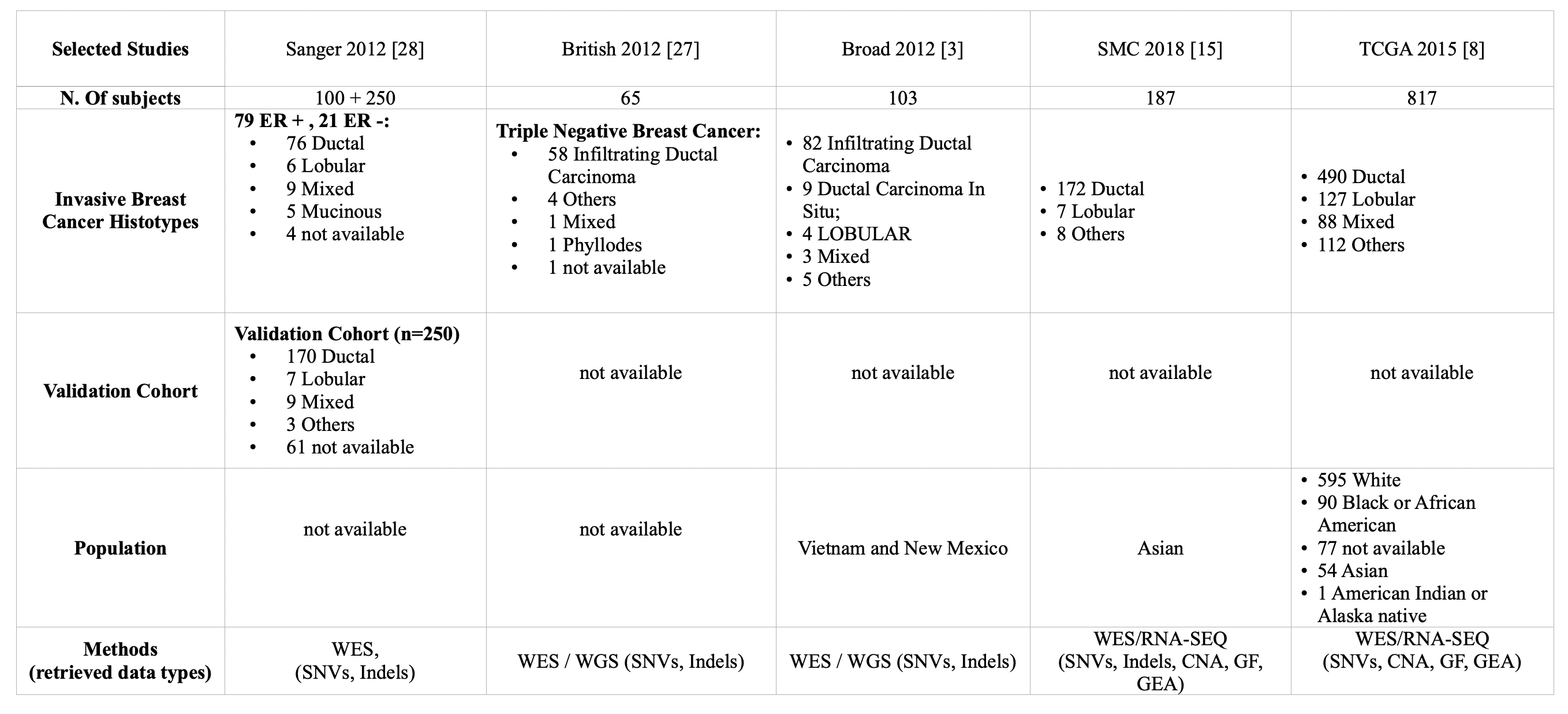}
\caption{\textbf{Breast cancer studies selected from the cBioPortal for Cancer Genomics}. The table reports some details on the invasive Breast Cancer populations used in the experiments. A study was selected if it included human participants AND at least a whole exome sequencing analysis was performed (studies performing targeted sequencing analyses were ruled out) ER: Estrogen receptor, WES: Whole Exome Sequencing, WGS: Whole Genome Sequencing, SNVs: Single Neucleotide Variants, CNA: Copy Number Alteration, GF: Gene Fusions, GEA: Gene Expression Alteration}
\label{fig:population_study_table}
\end{figure}

\section{Methods}

\subsection{Gene co-expression networks}\label{subse:gcn}
At a high level, a Gene Co-expression Network (GCN) is an undirected 
graph, with genes as the vertices. An undirected edge exists between two 
nodes $i$ and $j$ if the corresponding genes are significantly 
co-expressed in the subject population.\footnote{In the remainder we 
use the same letter to denote a gene and the corresponding vertex in 
the PPI or GCN.} Hence, differently from a PPI, 
a GCN is disease-specific. Intuitively, for two genes 
$i$ and $v$, this means that the transcript levels of $u$ and 
$j$ tend to vary in a similar fashion across the subject population.
Gene co-expression networks \cite{10.1093/bioinformatics/bth234} are of biological interest since 
co-expressed genes tend to be controlled by the same transcriptional 
regulatory program, i.e., they are functionally related or members of the same 
pathway or protein complex.

\subsection{Biological Random Walks}

\subsubsection{Overview}
As we decribed in the main article, we 
compute the personalization vector $\mathbf{q}$ and the transition matrix 
$W$ by leveraging multiple biological information sources. We remind that we refer to the biological information associated to a gene \textit{i} (e.g., the set of its annotations) as the set of its \textit{features}, denoted by \textit{features(i)}, such as the Gene Ontology database 
\cite{DBLP:journals/nar/Consortium19a} (GO in the remainder) or gene 
expression levels and their consequently differential expression scores. BRW ranks genes according to three main steps:

\begin{enumerate}
\item Rather than using the standard 
approach\footnote{We remind that in the standard RWR approach 
\cite{kohler2008walking}, the probability of restarting the random walk 
from a given seed node (disease gene) is the same for all seeds nodes, 
while it is $0$ for other nodes of the PPI.}, we compute a 
personalization vector $\mathbf{q}$, \textit{i.e} \textit{Biological Teleporting Probability} (BTP), associating individual 
teleporting probabilities to all nodes of the PPI. Qualitatively 
speaking, the probability of teleporting to a given node increases, the higher its 
feature-based similarity with the disease.
\item In a similar fashion, we derive the transition matrix $W$ of the 
random walk by also leveraging biological information, whereby the 
probability of a transition between genes $i$ and $j$ tends to be 
higher, the more functionally closer the genes are.
\item Finally, we rank genes according to their 
\textit{Biological Random Walk (BRW)} scores.
\end{enumerate}

\subsection{Biological Teleporting Probability}\label{subse:teleport}
Both annotation data and gene expression levels allow derivation of 
personalization vectors that reflect some notion of similarity between a gene and a disease, \textit{i.e.} its seed genes. In the first case, this similarity is defined in 
terms of knowledge about genes' involvement in various biological 
functions and/or diseases. In the second case, similarity is defined in 
terms of co-expression levels of different genes in subject cases as 
opposed to expression levels in a control group, using data about a 
population of patients involved in a clinical trial.
It's worth noting that this approach allows to easily extend the algorithm on a multiomics approach and/or tissue specific analysis, as long as it is possible to recapitulate data as a score for each gene (\textit{i.e.}chromatin accessibility) or pair of genes (\textit{i.e.} co-targeted by miRNA).

\begin{figure}[h]
\centering
\includegraphics[width=0.8\textwidth]{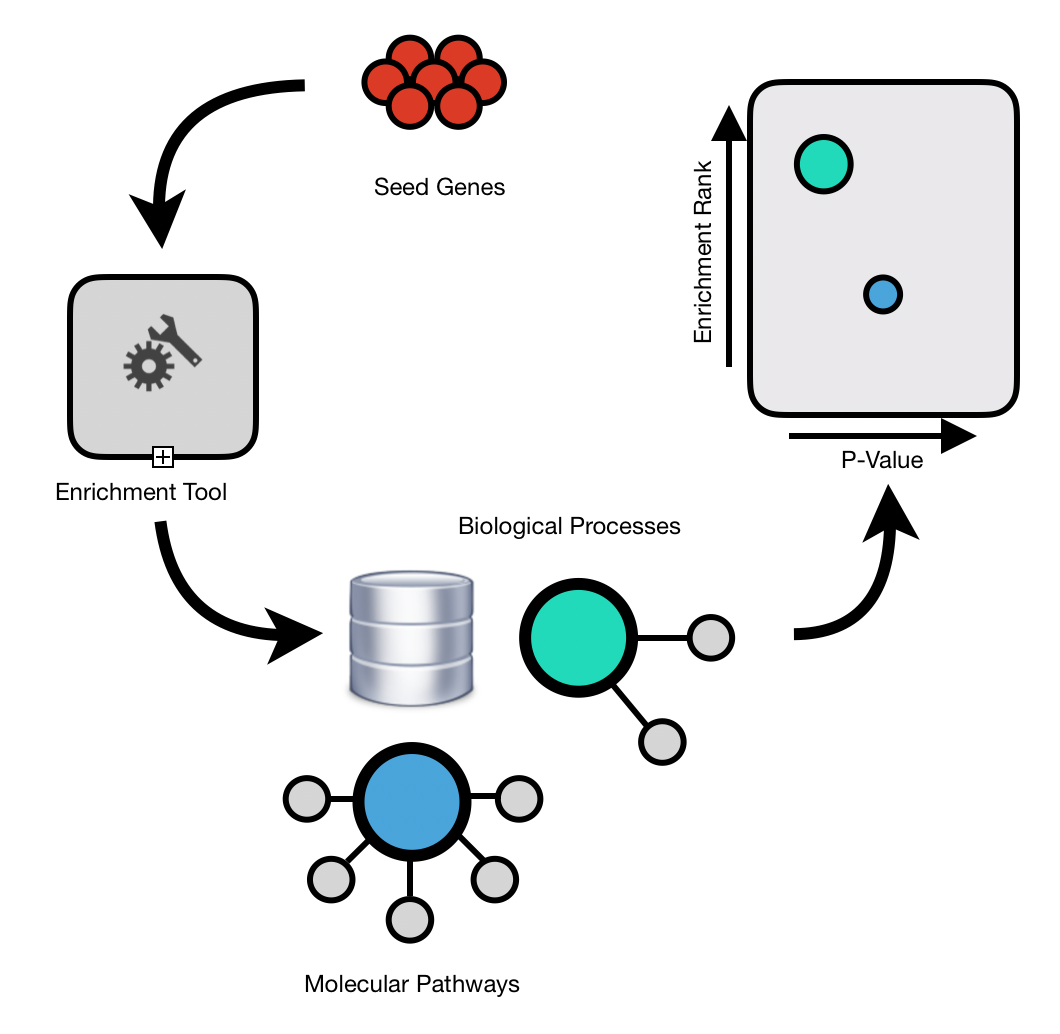}
\caption{Enrichment step: starting from a seed set $S$, We filter out $S$' annotations that are not statistically significant using Fisher Exact Test and FDR correction. }
\label{fig:enrichment_step}
\end{figure}

\subsubsection{Computing a Personalization Vector}\label{subsubse:pv}
Building on the hypothesis that genes involved in the same disease 
tend to be functionally related and thus share similar biological 
information, a first approach to deriving the personalization vector 
$\pv$ relies on assigning scores to genes in the PPI 
using biological information.

\paragraph{Using Biological Information.}
We assume we have $\ell$ sources of biological information. Assume $S$ 
is the seed set. for every $j=1,\ldots, \ell$, we use $\source^j$ to 
denote the subset of annotations from the $j$-th source that are 
associated with at least one gene in $S$. Then, for every $j=1,\ldots, 
\ell$, we selected the subset of annotation $\widehat{\source^j}$ 
filtering out the annotations that are not statistically significant as 
shown in Supplementary Fig.\ref{fig:enrichment_step} (i.e., $p-value > 10^-5$ using 
Fisher Exact Test and FDR correction) so that $\source = 
\cup_{j=1}^{\ell}\widehat{\source^j}$ denotes the set of all 
statistically significant annotations that are associated with genes in 
$S$. Likewise, for every gene $i$ (not necessarily belonging to $S$), 
we denote by $\ann(i)$ the set of its annotations, possibly extracted 
from multiple biological information sources.  We next assign each gene 
$i$ a weight $\weight_i$ as follows: 

\begin{equation}
	\weight_i =  
	\begin{cases}
		\ell & \text{if } i\in S,\\
		\sum_{j=1}^{\ell} \frac{|\ann(i)\cap \source^j|}{|\source^j|} & 
		\text{otherwise},
	\end{cases}
\end{equation}

\noindent if we want to weigh the seed set $S$ more than the subset 
of genes that are biologically similar to $S$. Otherwise, we can assign 
weights as follows:

\begin{equation}
	\weight_i =  
		\sum_{j=1}^{\ell} \frac{|\ann(i)\cap \source^j|}{|\source^j|} 
\end{equation}

If no biological information is used (i.e., we are only using knowledge 
of whether a given gene belongs to the seed set or not like in 
\cite{kohler2008walking}), each gene $i$ is assigned a weight 
$\weight_i$ as follows:
\begin{equation}
	\theta_i = 
	\begin{cases}
		1 & \text{if } i \in S \\
		0        & otherwise
	\end{cases}
\end{equation}

At this point, components of a personalization vector $\pv$ can be computed as follows:
\begin{equation}
q_i = \frac{\theta_i}{\sum_{i = 0}^{N} \theta_i },
\end{equation}
with $N$ the number of genes we consider. Note that $q_i$ denotes the 
probability that, upon teleportation, the random walkers jumps to gene 
$i$.

\paragraph{Using gene expression data.}

RNA sequencing techniques have changed our ability to explore the 
molecular mechanisms underlying complex diseases and they have been 
used to identify potential disease-associated genome-wide changes in 
gene expression patients. An important goal is the identification of 
Differentially Expressed (DE) genes, whose expression levels
systematically differ between case (Breast Cancer tissues) and a 
control group (Healthy Breast tissues).

We take inspiration from the framework to construct and 
integrate personalized perturbation profiles (PEEPs) from gene 
expression data proposed in \cite{menche2017integrating}, to derive 
gene expression-aware personalization vectors.

PEEPs turn individual expression heterogeneity into a 
predictive information by constructing personalized perturbation 
profiles that reflect expression changes within a single subject. For 
each gene $i$ and for each subject $j$, the expression 
level $l_i^j$ is compared with the distribution of the expression level 
of gene $i$ within the control group. This comparison is 
measured by the corresponding z-score, defined as:
\begin{equation}
 z_{ij} = \frac{l_i^j - \mu_i}{\sigma_i},
\end{equation}

where $\mu_i$ and $\sigma_i$ respectively denote the mean and the standard 
deviation of gene $i$'s expression level in the control group. This 
approach allows association of a ``bar code'' 
\cite{menche2017integrating} to each subject, thus allowing to tell
differentially expressed genes ($|z_{ij}| > 2.5$) from regular ones.

In general, given a case and a control group, it is possible to build a matrix $Z \in R^{n\times m}$, where $n$ and $m$ are respectively the number of genes and subjects:
\begin{equation}
Z = \begin{bmatrix}
z_{11}&z_{12}&\cdots &z_{1m} \\
z_{21}&z_{22}&\cdots &z_{2m} \\
\vdots & \vdots & \ddots & \vdots\\
z_{n1}&z_{n2}&\cdots &z_{nm}
\end{bmatrix}
\end{equation}

Thus, we might consider only one gene and subject at time, or 
we might consider a gene $i$ as the vector $Z_i$  
of the $z$-scores of its expression levels in the subject group. This 
provides an overall description of how gene $i$'s expression varies within 
the case group.

Using matrix $Z$, we identify the subset of genes that are 
differentially expressed in the subject group as follows. In 
particular, we define a new binary matrix $\hat{Z}$ which, for each 
gene $i$ and patient $j$, specifies whether or not $i$ is 
differentially expressed in subject $j$. Following 
\cite{menche2017integrating}, we assume this to happen when its 
$z$-score $z_{ij}$ exceeds $2.5$ standard deviations. Namely:

\begin{equation}
    \hat{Z}_{ij} = 
     \begin{cases}
     1 & \text{if } |z_{ij}| > 2.5  \\
      0        & otherwise
        
     \end{cases}
\end{equation}

Next, a gene $g$ is considered to be \emph{differentially expressed} 
in the subject group if the number of patients in which it is 
differentially expressed according to the above definition exceeds the 
mean:	
\begin{equation}
    \sum_{j} \hat{Z}_{gj} > \frac{1}{n}\sum_{i}\sum_{j} \hat{Z_i^j}.
\end{equation}

\medskip\noindent
The subset of differentially expressed genes this identified can be 
used to define a personalization vector $\pv$, using the PPI to identify 
differentially expressed genes that are also close to known disease 
genes (seeds) in the PPI. In particular, we obtain a personalization 
vector according to the following criteria: i) $\pv_i$ is non-zero only 
for differentially expressed genes; ii) $\pv_i$ increases with the 
number of disease genes in the vicinity of $i$.
In particular, for a differentially expressed gene $i$, we define a 
score $\phi_i$ as the proportion of its neighbors within shortest path 
distance $2$ that are disease proteins \cite{10.1093/bioinformatics/btl467}, namely:

\begin{equation}
    \phi_i = \frac{|N(i)\cap S|}{N(i)} +  \frac{|N_2(i)\cap S|}{N_2(i)}
\end{equation}
where we remind that $S$ denotes the seed set in the random walk with 
restart.
Finally, scores are normalized to obtain restart probabilities, so that:

 \begin{equation}
    \pv_i = \frac{\phi_i}{\sum_{i = 0}^{N} \phi_i } 
\end{equation}

\subsubsection{Integrating Biological Information and Gene 
Expression}\label{subsubse:int_pv}
We discussed above two orthogonal approaches to the design of personalization 
vectors. The first one leverages similarities between the biological 
processes associated to known disease proteins and those to be 
prioritized. Hence, teleporting probabilities depend on the seed set 
through association with common biological processes.

In the second case, teleporting probabilities depend on information 
that is tissue-specific (the level of expression in a population of 
subjects affected by a certain disease) and partly on the seed set, but 
this time through the PPI's topology. Hence, these two approaches 
largely rely on complementary sources of information. In the following 
paragraphs, we discuss ways to integrate these complementary sources 
into a unique personalization vector that leverages both.

\medskip\noindent
Assume we have computed two personalization vectors $\pv_1$ and 
$\pv_2$, the former using biological information only, the latter using 
gene expression data and the PPI. 

A first approach to integrating them is considering a convex 
combination, namely:
\begin{equation}
	\pv = \alpha\pv_1 + (1 - \alpha)\pv_2,
\end{equation}

where $\alpha\in [0, 1]$. Parameter $\alpha$ allows to weigh the 
importance of the different information sources we are using. In our 
experiments, we set $\alpha = 1/2$. Intuitively, this type of 
aggregation amounts to considering a gene a potential candidate if it 
is statistically significant in terms of its involvement in biological 
processes, gene expression levels in the subject group, or both.

\noindent\textbf{Remark.} Despite its simplicity, this is a 
mathematically principled choice. In particular, it is well-known \cite{jeh2003scaling} and easy to show that the
stationary distribution corresponding to the convex combination of two personalization vectors $\pv_1$ and $\pv_2$ is itself the linear combination of the stationary
distributions corresponding to $\pv_1$ and $\pv_2$ respectively. Briefly put, the parameter $\alpha$ allows us to tune the relative importance of the information provided
by gene annotations and gene expression levels respectively. Moreover, this approach extends seamlessly to an arbitrary number of heterogeneous sources of information
(and corresponding personalization vectors).

Alternatively, we can consider a gene to be a potential candidate if it 
is statistically significant both in terms of the biological processes 
it is involved in (with respect to the seed of known disease genes) and 
of its expression levels in the population of subjects. In this case, a 
natural way to define $\pv$ is as follows:

\begin{equation}
    \pv_i = 
    \frac{\pv_{1i}\cdot\pv_{2i}}{\sum_{j=1}^n\pv_{1j}\cdot\pv_{2j}}.
\end{equation}
Note that, again, $\pv$ is a probability distribution over the genes.

\subsection{Biological Transition Matrix}\label{subse:trans_mat}

\subsubsection{Computing a Transition Matrix}
Similar approaches can be used to derive a transition matrix for the 
random walk with restart. Both approaches rely on the PPI, differing on 
the way PPI's edges are assigned weights that reflect genes' similarity 
and determine the probabilties of edge traversals. 

\subsubsection{Using Biological Information}
In this case, we consider a weighted transition matrix $\tmat$, in 
which each entry $W_{ij}$ depends on the extent to which nodes genes 
$i$ and $j$ of the PPI share common annotations (i.e., they are involved in common 
biological processes) that are also significant for the disease. In 
more detail, considered genes $i$ and $j$, we define the following 
\emph{Disease Specific Interaction Function} (DSI function in the remainder):

\begin{equation}
    \dsi(i,j) = |\ann(i)\cap\ann(j)\cap\source|,
\end{equation}
where we remind that $\ann(k)$ denotes the set of gene $k$'s 
annotations, while $\source$ denotes the overall set of annotations 
that are statistically significant for disease genes.

Intuitively, the higher $\dsi(i,j)$, the more $i$ and $j$ 
share annotations that are also statistically significant for the 
disease under consideration.

\noindent In the end, edges in the PPI will be assigned weights 
depending on the DIS function as follows:
\begin{equation}
    \tmat_{ij} =
    \begin{cases}
     c + \dsi(i,j) & \text{if } (i,j) \in E, \\
      0        & otherwise.
    \end{cases}
  \end{equation}
Here, $c$ is a positive constant that accounts for usual sparsity of 
the available datasets, so that no biological information may be 
available for the end-points of a link in the PPI. In this case, the 
link receives a minimum weight $c$.

\subsubsection{Using Co-Expression Information}

In this case, we use gene expression information about a population of 
patients, in order to augment the PPI with tissue-specific information. 
Specifically, gene expression information is used to assign weights to 
edges, this time reflecting similarities between genes in terms of 
co-expression with respect to the subject population.

\noindent Consider a CO-Expression network in which each pair of gene $(i,j)$ is associated a score defined by the Pearson's correlation coefficient $pc_{i,j}$. We need $pc_{i,j}$ to be a probability distribution on PPI network for each gene $i \in V$. Thus, we define the probability of the random walker to go from node \textit{i} to one of its neighbors 
\textit{j} in the PPI network as:
\begin{equation}
    \tmat_{ij} = \frac{pc_{ij}}{\sum_{k \in \neigh(i)} pc_{ik}}
\end{equation}

\noindent where \textit{N(i)} is the set of \textit{i's} neighbors in the PPI network.

\subsubsection{Computing an Aggregate Transition 
Matrix}\label{subse:mat_agg}
In the previous sections, we have seen how we can derive random walk's 
transition matrices using only information about biological processes 
(i.e., annotations) or gene expression data from a population of 
subjects. While similar approaches apply to other prospective 
information sources, we show below how 
transition matrices derived from the aforementioned sources can be 
integrated into an \emph{aggregate transition matrix}, leveraging both 
information sources.

\smallskip
Assume we have computed transition matrices $\tmat_1$ and $\tmat_2$ 
using biological and gene expression information respectively. We 
proceed in a way that is similar to the approach used to derive an 
aggregate personalization vector.

\smallskip
A first approach is considering a convex combination. Namely, for 
$\beta\in [0, 1]$, we shall consider the transition matrix
\[
	\tmat = \beta\tmat_1 + (1 - \beta)\tmat_2.
\]
Note that choosing $\beta = 1$ corresponds to only considering 
biological annotations, whereas $\beta = 0$ corresponds to only 
leveraging gene expression data. So, $\beta$ is a parameter, whose 
tuning allows us to weigh the importance of one source of information 
with respect to the other. This approach can be extended to the case of 
more than two information sources in the obvious way.

\subsection{Network Randomization}\label{subse:apx_random}
For the sake of completeness, in this section, we succintly describe 
the degree-preserving, network randomization algorithm proposed in
\cite{milo2003uniform}.
Given an initial network $G(V,E)$, the algorithm involves carrying out 
a series of Monte Carlo switching steps whereby a pair of edges $(A 
\rightarrow B, C \rightarrow D)$ is selected at random and the ends are 
exchanged to give $(A \rightarrow D, C \rightarrow B)$.

\begin{figure}[h]
\centering
\includegraphics[width=0.6\textwidth]{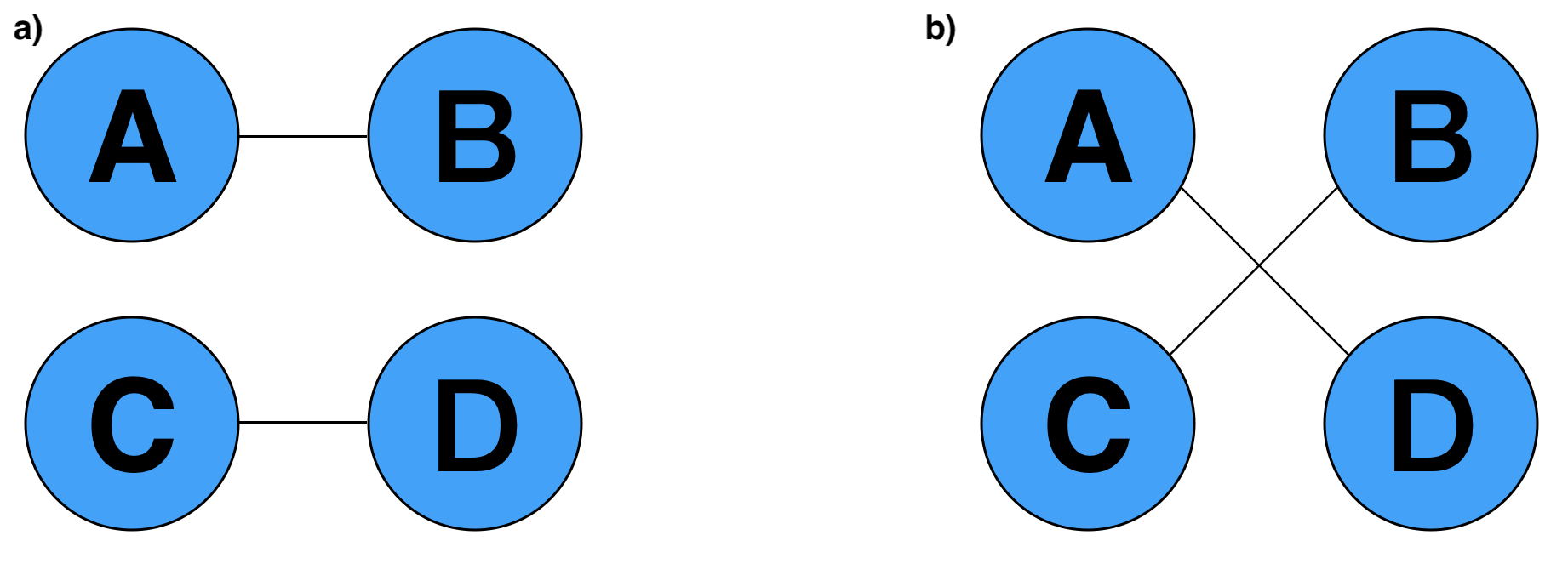}
\caption{Randomization process: \textbf{a)} shows the original edges in the network. \textbf{b)} shows the edges computed after the switching phase}
\label{fig:network_randomization_how_it_works}
\end{figure}

Supplementary Fig. \ref{fig:network_randomization_how_it_works} shows the original edges and the switched edges after one step of the Markov Chain Monte-Carlo Switching Algorithm. However, the switching phase is not performed in two different cases:

\begin{itemize}
    \item If it generates multiple edges. For instance if it chooses the pair of edges $(A \rightarrow D, C \rightarrow D)$ a duplicate edges is created as shown in figure \ref{fig:network_randomization_duplicate_edges}.
    
    \begin{figure}[h]
    \centering
    \includegraphics[width=0.6\textwidth]{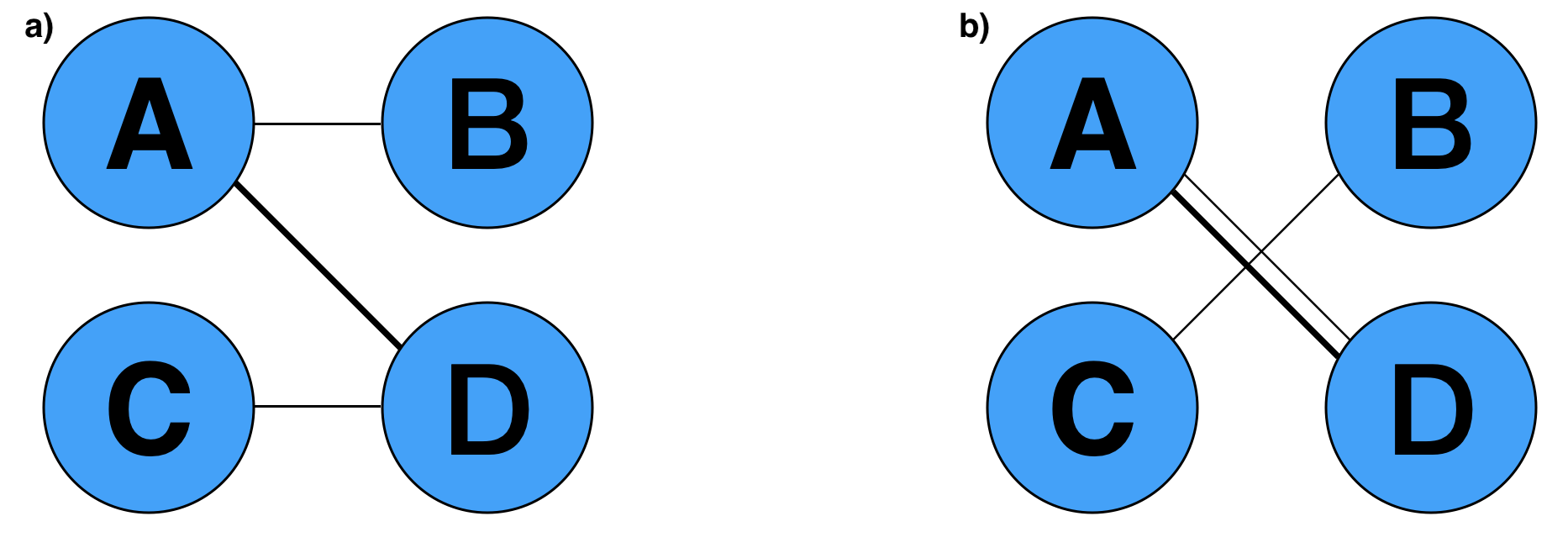}
    \caption{Multiple edge generation: \textbf{a)} shows the original edges in the network. \textbf{b)} shows the duplicated edge computed after the switching phase}
    \label{fig:network_randomization_duplicate_edges}
    \end{figure}
    
    \item If it generates self-edges. If it randomly select the pair of edges $(B \rightarrow A, A \rightarrow C)$, a self-loop is created as show in figure 
    
    \begin{figure}[h]
    \centering
    \includegraphics[width=0.6\textwidth]{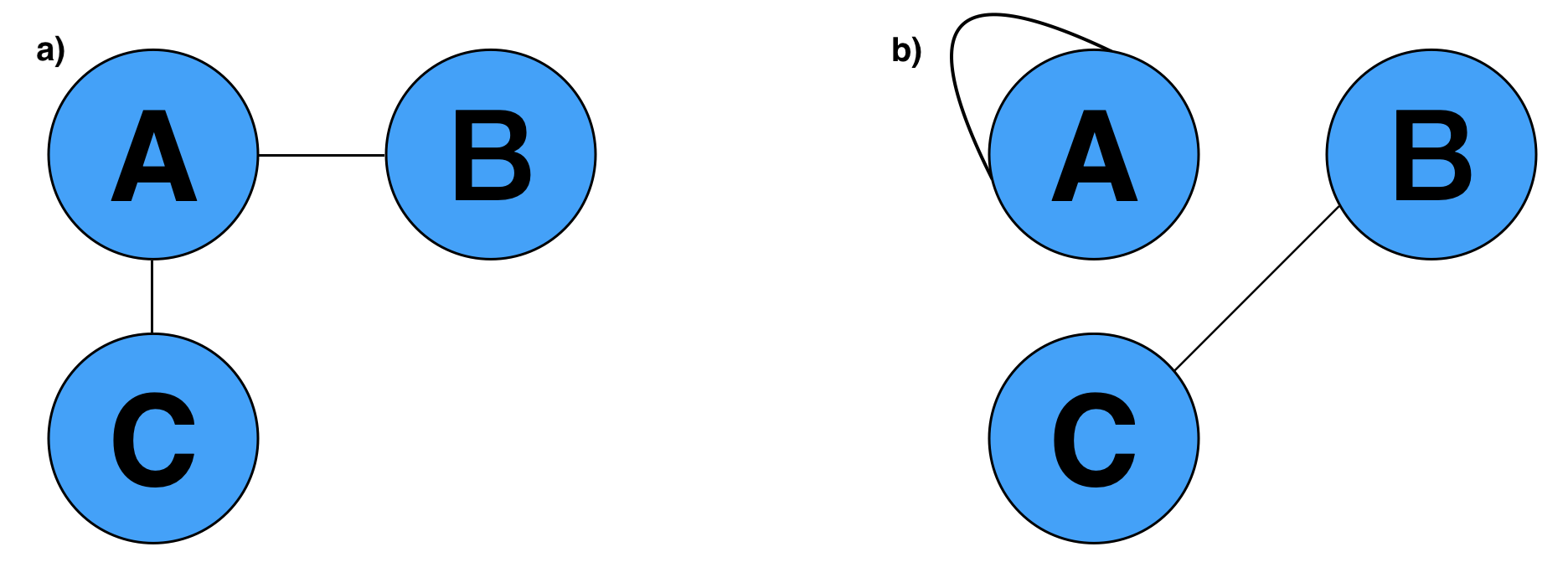}
    \caption{Multiple edge generation: \textbf{a)} shows the original edges in the network. \textbf{b)} shows the self-loop computed after the switching phase}
    \label{fig:network_randomization_self_loop}
    \end{figure}
\end{itemize}

\noindent The entire process is repeated exactly $Q \cdot |E|$ times, where $E$ is the number of edges in the original network and $Q$ is chosen large enough to have a good mixing.

\section{Results}\label{apx:results}

\subsection{Performance indices}\label{apx:evaluation_scheme}
To assess accuracy in disease gene prioritization, for each disease analyzed, we used a 100-fold random validation approach, with  each  random  split  consisting of a training set of seed genes (corresponding to 70\% of the total) that were provided as input to the algorithms and a test set, accounting for the remaining 30\% of the known disease genes . To evaluate the power of each heuristic, we considered two standard prediction indices in Data Mining:
\begin{itemize}
    \item \textbf{Recall@$k$:} this is the fraction of the test set 
    that is successfully retrieved in the top $k$ positions of the ranking computed by the algorithm. 
    
    \item \textbf{nDCG:} The normalized Discounted Cumulative Gain (nDCG) is proven to be able to select the better ranking between any two, substantially different rankings. For binary classification the nDCG is given by:
    \begin{center}
        $\frac{\sum_{i \in P} \frac{1}{log_2(i + 1)}}{\sum_{i = 0}^{|P|} \frac{1}{log_2(i + 1)}}$
    \end{center}
    
Where summation in the numerator runs over all positive instances, while summation in the denominator quantifies the ideal case, in which positive instances appear in the top positions of the algorithm's ranked list.
\end{itemize}

\subsection{Multi-Omics Integration}\label{subse:integration}
\begin{figure}[h]
\centering
\includegraphics[width=\textwidth]{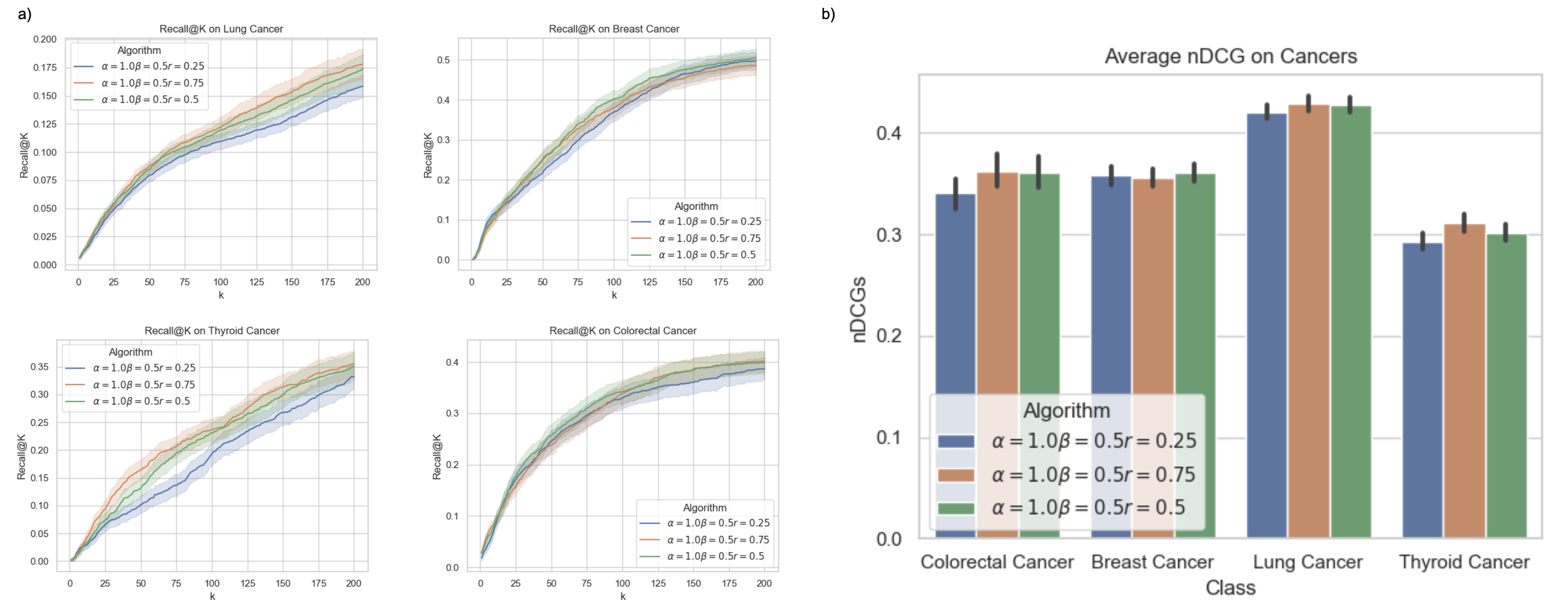}
\caption{Benchmark on each cancer disease analyzed to pick the best 
combination of hyper parameters $(\alpha,\beta,r)$. Tab \textbf{a)} shows 
the ccomparison with respect to Recall@k and Tab \textbf{b)} 
comparison with respect to nDCG, for different values of the 
teleporting probability. Regardless of the choice of $r$, results (in 
terms of Recall@k and nDCG) were best when $\alpha = 1$ and $\beta = 
0.5$. These values correspond to i) considering only ontologies to the 
purpose of computing personalization vectors and ii) giving equal 
importance to ontologies and gene expression when computing the 
transition matrix of the random walk.}
\label{fig:benchmarking_restart}
\end{figure}

\noindent To identify the best combination of hyper-parameters, we 
tested how different integration approaches affect BRW's  
performances on the same validation 
benchmark discussed in the main article, in terms of Recall@k and nDCG. 
In more detail, we 
considered 3 different validation approaches: i) Monte Carlo Random 
Sampling, ii) Drug Target Discovery and iii) Drug Enrichment. 

\paragraph{Monte Carlo Sampling.} We considered all cancer 
phenotypes analyzed in the main article, and for each phenotype we 
considered different values of $\alpha$ and $\beta$, fixing the restart 
probability $r$ to 0.5 in order to reduce grid search space. 
We found out that manually curated disease 
genes have a bias on the ontologies (GO biological processes, KEGG and 
Reactome pathways). Indeed, the best performance are obtained for 
large values of $\alpha$ as shown in Supplementary Fig. 
\ref{fig:benchmarking_alpha_beta}. I.e., it is 
better to teleport to biologically similar proteins than differentially 
expressed genes. As regards the aggregated transition matrix, 
performance in terms of Recall@k/nDCG is maximized when $\beta = 0.5$. 
In other words, performance is maximized when co-expression information 
is combined with biological information (and the PPI). Once we 
discovered the best combination of $(\alpha,\beta)$, we tried different 
values of the restart probability, with results summarized in Supplementary Fig. 
\ref{fig:benchmarking_restart}. As Supplementary Fig. 
\ref{fig:benchmarking_restart} shows, results are not greatly affected by the 
value of the restart probability, as long this as this exceeds the 
value $0.25$.

\paragraph{Drug target discovery and drug enrichment.} In this case, we considered 
Breast Cancer phenotype and we analyzed drug enrichment and performance in 
drug target discovery of each combination of hyper-parameters $(\alpha, 
\beta, r)$, in terms of Recall@k and nDCG. Table \ref{tab:drug_enrichment_benchmark_best_combination} 
shows the result of the drug enrichment, Recall@k, nDCGs and number of 
FDA approved, Breast Cancer drugs that were annotated, for each 
combination of parameters. As shown in the table, different parameter 
combinations perform best for different validation tasks. For example, 
parameter combinations corresponding to a bias toward ontologies 
perform best in the task of identifying Breast Cancer drug targets of 
approved drugs, highly scoring drug targets in the first 200 positions, 
whereas biasing the random walk towards differentially 
expressed genes and the co-expression network performs best in prioritizing genes that are enriched 
in many relevant drugs. Indeed, Paclitaxel, Fulvestrant, 
Doxorubicin,Tamoxifen, Methotrexate, and letrozole, well-known approved Breast Cancer drugs, obtain are significantly enriched when gene expression is weighted more than ontologies. Possible explanations for this phenomenon are 
given in Sections \ref{subsub:target} and \ref{se:concl}.

\begin{figure}[p]
\centering
\includegraphics[width=0.9\textwidth]{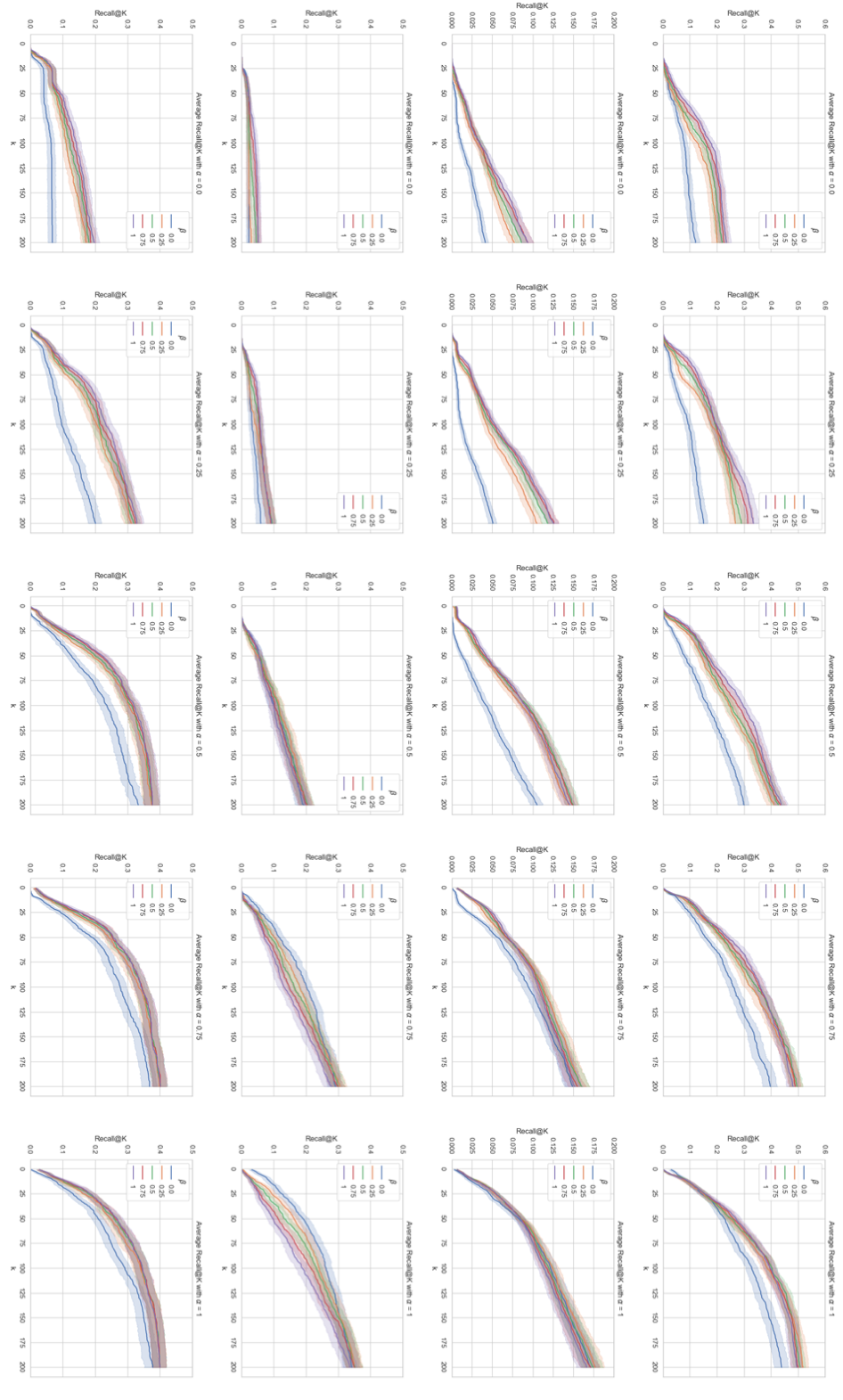}
\caption{Benchmark on each cancer disease analyzed to identify the best 
combination of pair of hyper parameters $(\alpha,\beta)$. Each row 
represents the following tumor type analyzed in this manuscript: BRCA, 
LUAD, THCA, and COAD. Each row plots the comparison of different runs 
of the BRW for same value of $\alpha$ and different values of $\beta$, 
while restart probability is set to 0.5 in all experiments.}
\label{fig:benchmarking_alpha_beta}
\end{figure}

\begin{table}
\resizebox{\textwidth}{!}{%
\begin{tabular}[!htp]{lp{0.5cm}p{0.5cm}p{0.5cm}p{0.5cm}p{0.5cm}p{0.5cm}p{0.5cm}p{0.5cm}p{0.5cm}p{0.5cm}p{0.5cm}p{0.5cm}p{0.5cm}p{0.5cm}p{0.5cm}p{0.5cm}p{0.5cm}p{0.5cm}p{0.5cm}p{0.5cm}p{0.5cm}p{0.5cm}p{0.5cm}p{0.5cm}p{0.5cm}p{0.5cm}p{0.5cm}}

Hyper Parameter Combination ($\alpha$, $\beta$, $r$) & (0.25, 0.25, 0.25) & (0.25, 0.25, 0.5) &  (0.25, 0.25, 0.75) &  (0.25, 0.5, 0.25) &  (0.25, 0.5, 0.5) &  (0.25, 0.5, 0.75) &  (0.25, 0.75, 0.25) &  (0.25, 0.75, 0.5) &  (0.25, 0.75, 0.75) & (0.5, 0.25, 0.25) &  (0.5, 0.25, 0.5) &  (0.5, 0.25, 0.75) &  (0.5, 0.5, 0.25) & (0.5, 0.5, 0.5) &  (0.5, 0.5, 0.75) &  (0.5, 0.75, 0.25) &  (0.5, 0.75, 0.5) &  (0.5, 0.75, 0.75) &  (0.75, 0.25, 0.25) & (0.75, 0.25, 0.5) & (0.75, 0.25, 0.75) & (0.75, 0.5, 0.25) &  (0.75, 0.5, 0.5) &  (0.75, 0.5, 0.75) & (0.75, 0.75, 0.25) &  (0.75, 0.75, 0.5) &  (0.75, 0.75, 0.75) \\
estradiol        &      \cellcolor[HTML]{A8A8A8}\textbf{18.7} &              10.8 &                10.1 &               11.9 &               7.4 &                8.8 &                10.1 &                7.4 &                 8.1 &              11.8 &               7.9 &                5.9 &              12.2 &             7.0 &               6.2 &               10.9 &               7.1 &                6.2 &                 8.5 &               4.2 &                3.3 &              11.5 &               4.2 &                3.7 &               11.6 &                5.2 &                 3.7 \\
imatinib         &      \cellcolor[HTML]{A8A8A8}\textbf{15.5} &               6.9 &                 5.9 &               11.0 &               6.4 &                5.8 &                 8.3 &                7.1 &                 5.5 &              13.5 &               7.9 &                5.9 &              10.6 &             7.4 &               6.9 &                8.6 &               7.4 &                7.1 &                10.5 &               7.8 &                7.1 &              10.4 &               8.0 &                6.1 &               10.3 &                7.3 &                 6.4 \\
arsenic trioxide &                9.5 &               6.3 &                 5.1 &               11.0 &               5.8 &                5.0 &                10.3 &                7.1 &                 5.3 &              15.6 &               9.8 &                5.5 &              15.0 &            10.8 &               6.9 &               15.3 &              12.3 &                6.5 &                15.6 &              10.8 &               11.8 &     \cellcolor[HTML]{A8A8A8}\textbf{17.7} &              13.9 &               10.8 &               17.1 &               13.9 &                11.3 \\
bexarotene       &                7.2 &               4.6 &                 4.8 &                9.5 &               5.2 &                4.1 &                 9.8 &                5.7 &                 3.7 &               5.7 &               6.7 &                4.3 &               8.9 &             7.0 &               4.3 &                9.3 &               7.4 &                4.3 &                 5.2 &               3.7 &                3.6 &               8.0 &               5.2 &                4.6 &      \cellcolor[HTML]{A8A8A8}\textbf{11.0} &                6.5 &                 4.0 \\
\textbf{paclitaxel}       &                4.8 &      \cellcolor[HTML]{A8A8A8}\textbf{6.3} &                 4.6 &                4.7 &               5.0 &                4.5 &                 4.2 &                5.2 &                 4.2 &               3.4 &               4.2 &                4.7 &               3.8 &             3.9 &               4.7 &                3.5 &               4.2 &                4.5 &                 3.0 &               2.9 &                2.7 &               4.0 &               2.5 &                2.4 &                4.6 &                3.1 &                 2.5 \\
sirolimus        &               10.1 &               4.5 &                 4.3 &               11.0 &               4.6 &                4.2 &                10.1 &                4.7 &                 4.6 &               8.9 &               7.0 &                4.3 &              13.4 &             6.6 &               3.8 &               15.3 &               7.1 &                3.9 &                12.2 &               7.8 &                5.0 &              17.7 &              11.3 &                4.9 &      \cellcolor[HTML]{A8A8A8}\textbf{19.1} &               13.4 &                 5.3 \\
\textbf{fulvestrant}      &       \cellcolor[HTML]{A8A8A8}\textbf{9.1} &               4.4 &                 4.2 &                5.7 &               2.6 &                4.1 &                 4.0 &                2.9 &                 3.8 &               5.7 &               3.1 &                2.0 &               4.7 &             2.2 &               2.1 &                2.9 &               2.0 &                2.3 &                 3.0 &               0.9 &                1.1 &               2.5 &               1.1 &                0.9 &                2.4 &                1.1 &                 1.0 \\
valproic acid    &       \cellcolor[HTML]{A8A8A8}\textbf{8.5} &               3.7 &                 3.5 &                7.3 &               3.2 &                3.5 &                 4.4 &                3.3 &                 3.4 &               5.7 &               4.2 &                3.1 &               5.1 &             3.6 &               3.1 &                5.8 &               3.1 &                3.1 &                 3.8 &               3.0 &                2.4 &               5.1 &               3.8 &                2.4 &                5.9 &                4.7 &                 2.8 \\
vitamin c        &       \cellcolor[HTML]{A8A8A8}\textbf{9.1} &               4.4 &                 3.5 &                5.5 &               3.4 &                3.2 &                 6.6 &                3.3 &                 2.9 &               7.0 &               4.5 &                2.0 &               6.6 &             3.6 &               2.0 &                6.1 &               3.3 &                2.1 &                 6.9 &               3.3 &                2.4 &               6.6 &               4.7 &                2.5 &                8.8 &                5.3 &                 2.5 \\
\textbf{doxorubicin}      &       \cellcolor[HTML]{A8A8A8}\textbf{7.6} &               4.1 &                 3.5 &                5.7 &               3.6 &                3.6 &                 6.6 &                4.4 &                 3.8 &               5.7 &               5.8 &                4.5 &               6.6 &             4.9 &               4.1 &                6.5 &               5.3 &                3.9 &                 5.1 &               5.6 &                4.8 &               7.6 &               6.6 &                4.6 &                7.6 &                6.7 &                 4.6 \\
\textbf{tamoxifen}        &      \cellcolor[HTML]{A8A8A8}\textbf{12.4} &               5.3 &                 3.2 &                5.6 &               3.3 &                2.6 &                 6.3 &                4.0 &                 2.5 &               6.2 &               4.9 &                2.4 &               5.7 &             4.3 &               2.9 &                7.4 &               4.0 &                3.2 &                 5.5 &               3.5 &                3.7 &               5.7 &               3.5 &                3.8 &                6.4 &                4.0 &                 4.3 \\
\textbf{methotrexate}     &      \cellcolor[HTML]{A8A8A8}\textbf{14.2} &               6.5 &                 3.2 &               10.2 &               5.6 &                2.6 &                 6.8 &                6.0 &                 2.6 &               9.7 &               6.9 &                2.9 &               8.3 &             5.4 &               3.2 &                6.9 &               4.7 &                3.2 &                 6.8 &               3.0 &                2.2 &               7.0 &               3.0 &                1.8 &                6.3 &                3.6 &                 1.8 \\
\textbf{letrozole}        &       \cellcolor[HTML]{A8A8A8}\textbf{8.7} &               4.9 &                 3.2 &                4.0 &               2.4 &                3.1 &                 2.6 &                2.9 &                 2.6 &               3.1 &               3.9 &                1.5 &               2.7 &             3.0 &               1.9 &                1.8 &               3.0 &                2.3 &                 3.0 &               1.6 &                2.0 &               3.1 &               1.8 &                2.0 &                2.3 &                2.3 &                 2.0 \\
etoposide        &       \cellcolor[HTML]{A8A8A8}\textbf{8.9} &               2.9 &                 2.8 &                5.1 &               2.6 &                2.8 &                 5.7 &                3.0 &                 3.1 &               6.7 &               3.6 &                2.0 &               6.2 &             2.7 &               2.0 &                6.3 &               2.7 &                2.1 &                 6.1 &               3.9 &                2.2 &               7.5 &               4.4 &                1.9 &                6.4 &                4.0 &                 1.9 \\
deferoxamine     &                4.5 &               2.8 &                 2.7 &                3.7 &               4.1 &                2.2 &                 4.4 &                4.2 &                 2.2 &      \cellcolor[HTML]{A8A8A8}\textbf{5.9} &               3.6 &                2.1 &               3.7 &             4.2 &               2.1 &                4.5 &               4.2 &                2.7 &                 4.3 &               4.2 &                2.2 &               3.8 &               4.2 &                2.8 &                5.4 &                5.1 &                 2.8 \\
vorinostat       &                7.7 &               4.0 &                 2.6 &                7.1 &               3.6 &                2.6 &                 8.4 &                4.7 &                 2.5 &               7.5 &               5.2 &                4.3 &               7.6 &             4.7 &               4.1 &                9.0 &               5.2 &                4.0 &                 5.2 &               6.0 &                4.2 &               7.6 &               6.9 &                4.6 &       \cellcolor[HTML]{A8A8A8}\textbf{9.1} &                7.3 &                 5.0 \\
cisplatin        &       \cellcolor[HTML]{A8A8A8}\textbf{7.4} &               3.9 &                 2.4 &                4.6 &               3.4 &                2.5 &                 4.7 &                3.5 &                 2.4 &               3.9 &               3.1 &                1.7 &               3.9 &             3.0 &               1.7 &                4.1 &               3.1 &                1.7 &                 4.3 &               3.7 &                2.6 &               4.0 &               3.1 &                2.6 &                6.0 &                4.0 &                 2.9 \\
cetuximab        &                2.8 &               2.8 &                 2.3 &                4.5 &               3.5 &                2.8 &                 4.3 &                3.8 &                 2.8 &               5.1 &               4.6 &                4.0 &               5.1 &             4.9 &               3.7 &                4.9 &               5.5 &                3.6 &                 5.3 &      \cellcolor[HTML]{A8A8A8}\textbf{7.2} &                6.2 &               5.8 &               6.6 &                6.2 &                5.6 &                6.3 &                 6.4 \\
zinc acetate     &                4.7 &               1.7 &                 2.0 &                4.7 &               2.1 &                2.0 &                 7.1 &                2.2 &                 2.0 &               4.1 &               3.6 &                1.6 &               7.8 &             3.5 &               1.6 &                7.2 &               4.0 &                1.6 &                 3.6 &               2.1 &                2.1 &      \cellcolor[HTML]{A8A8A8}\textbf{7.9} &               2.6 &                2.1 &                6.1 &                4.1 &                 2.1 \\
mifepristone     &                4.6 &               3.3 &                 1.7 &                5.0 &               3.7 &                1.7 &                 5.0 &                3.4 &                 1.7 &               3.9 &               3.8 &                2.0 &               5.0 &             4.2 &               1.7 &                4.5 &               4.3 &                1.7 &                 3.9 &               3.0 &                1.7 &               4.6 &               2.9 &                2.1 &       \cellcolor[HTML]{A8A8A8}\textbf{6.9} &                2.6 &                 2.1 \\
trovafloxacin    &                5.1 &               2.3 &                 1.6 &                3.4 &               2.5 &                1.4 &                 4.3 &                3.3 &                 1.1 &               5.7 &               4.0 &                2.0 &               5.1 &             3.6 &               2.0 &                4.9 &               4.1 &                2.0 &                 5.0 &               4.2 &                2.5 &               5.5 &               4.3 &                2.5 &       \cellcolor[HTML]{A8A8A8}\textbf{6.2} &                3.9 &                 2.5 \\
plicamycin       &                5.8 &               4.1 &                 1.6 &                5.7 &               3.9 &                1.8 &                 5.5 &                4.4 &                 1.8 &               5.4 &               4.7 &                2.9 &               4.9 &    \cellcolor[HTML]{A8A8A8}\textbf{6.6} &               3.2 &                4.2 &               6.4 &                3.4 &                 4.5 &               5.5 &                5.5 &               4.2 &               5.3 &                5.7 &                4.8 &                5.3 &                 5.7 \\
hydrocortisone   &                2.7 &               1.6 &                 1.3 &                4.8 &               2.1 &                1.8 &                 5.7 &                2.5 &                 2.1 &               4.0 &               3.3 &                2.2 &               5.7 &             4.2 &               2.6 &                6.2 &               4.2 &                2.6 &                 5.5 &               5.5 &                3.6 &               5.8 &               5.1 &                4.0 &       \cellcolor[HTML]{A8A8A8}\textbf{7.3} &                4.7 &                 4.1 \\
vincristine      &                2.0 &               1.1 &                 0.7 &                1.6 &               1.5 &                0.6 &                 2.4 &                1.5 &                 0.6 &               2.3 &               1.5 &                0.5 &               3.2 &             2.3 &               0.4 &                4.4 &               2.3 &                0.6 &                 2.3 &               1.5 &                0.6 &               4.4 &               1.8 &                0.6 &       \cellcolor[HTML]{A8A8A8}\textbf{6.3} &                2.3 &                 0.8 \\
hydroquinone     &                3.7 &               2.3 &                 0.6 &                3.3 &               2.7 &                0.8 &                 4.5 &                3.1 &                 0.8 &               4.0 &               3.2 &                2.2 &               3.7 &             3.9 &               2.2 &                4.5 &               3.5 &                2.6 &                 3.6 &               3.5 &                2.3 &               4.1 &               3.5 &                2.4 &       \cellcolor[HTML]{A8A8A8}\textbf{5.6} &                4.0 &                 2.4 \\
isotretinoin     &                1.5 &               0.9 &                 0.4 &                1.5 &               1.4 &                0.4 &                 2.1 &                1.4 &                 0.5 &               2.0 &               1.4 &                2.3 &               2.7 &             1.5 &               2.3 &                2.8 &               2.0 &                2.3 &                 4.1 &               4.7 &       \cellcolor[HTML]{A8A8A8}\textbf{5.3} &               3.5 &               4.0 &                5.3 &                2.8 &                4.1 &                 5.3 \\

Recall@200 &              0.16 &             0.09 &              0.07 &             0.19 &            0.11 &             0.09 &              0.19 &             0.12 &              0.09 &             0.19 &            0.16 &             0.12 &            0.19 &           0.19 &            0.14 &             0.23 &            0.19 &             0.14 &              0.21 &             0.25 &              0.21 &             0.23 &            0.26 &             0.23 &              0.25 &            \cellcolor[HTML]{A8A8A8}\textbf{0.26} &              0.23 \\
nDCG       &              0.35 &             0.33 &              0.32 &             0.37 &            0.34 &             0.32 &              0.40 &             0.34 &              0.32 &             0.39 &            0.36 &             0.34 &            0.40 &           0.37 &            0.34 &             0.41 &            0.37 &             0.34 &              0.39 &             0.40 &              0.37 &             0.40 &            0.40 &             0.37 &              0.41 &            \cellcolor[HTML]{A8A8A8}\textbf{0.41} &              0.38 \\
N. of Breast Cancer Annotated Drugs &15&9&6&17&9&7&17&12&7&17&15&6&17&17&9&17&17&9&17&17&13&17&17&15&17&17&15
\end{tabular}

}
\caption{This table describes drug enrichment and drug target discovery 
results for each combination of hyper-parameters $(\alpha,\beta, r)$. 
The table is divided into four sections: The first one (the header) 
represents all hyper-parameter combinations tested. Each combination is 
described by the corresponding triple $(\alpha, \beta, r)$. The second section 
shows the  $-log_{10}P_{value}$ of drugs enriched in at least one of 
the combinations. We considered only drugs with $P_{value} < 10^{-5}$ 
in at least one of the combinations. Drugs in bold are FDA-approved to 
treat breast cancer. The third section shows the portion of Breast 
cancer's drug targets predicted by each combination in the top 200 
positions (Recall@200) and the nDCG. Finally, the last section shows 
the number of Breast cancer's FDA-approved drugs with at least a drug 
target predicted in the first 200 positions by each combination of 
Hyper-Parameters.}

\label{tab:drug_enrichment_benchmark_best_combination}
\end{table}

\newpage

\subsection{Robustness and Bias}

\begin{figure}[h]
\centering
\includegraphics[width=\textwidth]{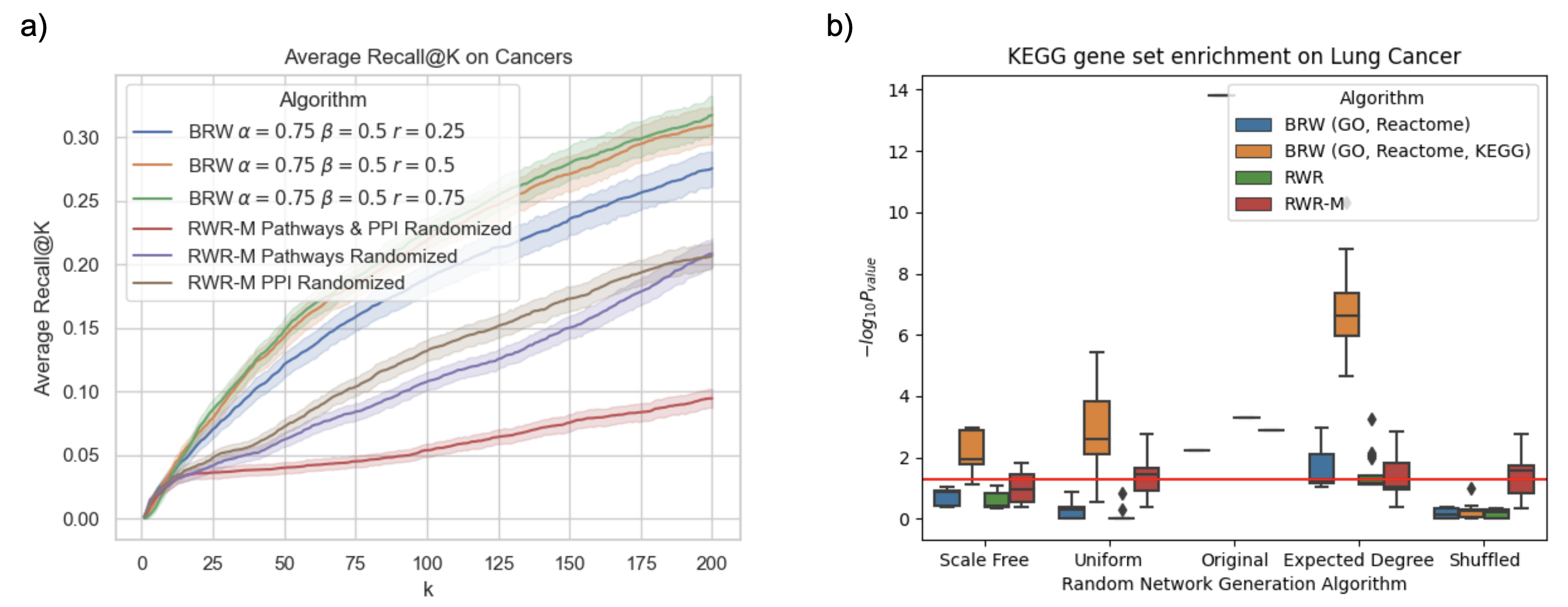}
\caption{BRW and RWR-M comparison. Tab a) shows the comparison between BRW 
and RWR-M on ranking test known disease genes when at least one network 
is randomized. As we can see, enriched ontologies (GO biological 
Processes, Pathways) correlate positively between seed and validation set. Consequently, the noise induced by the randomized PPI is mitigated. Unfortunately, even if RWR-M relies on several biological information (PPI- Network, CO-Expression Network and Pathways Network) it is not able to mitigate the noise as much is mitigated by
Biological Random Walks. Tab b) illustrates how multi-omics
integration affect the bias induced by curated ontologies (GO, 
Reactome) and the PPI network}.
\label{fig:robustness_and_bias}
\end{figure}

\noindent In a first experiment, we performed a mean 100-fold validation for each cancer 
phenotype by randomly sampling 70\% of known disease genes and using 
the rest to compute the average Recall@k of each tested algorithm. 
To begin, we tried different values of the restart probability to 
understand how the biologically weighted teleporting probability is 
correlated with the test genes. We found out that this correlation is 
strong, and thus when a very high restart probability is used, BRW 
improves its prediction performances, as shown in Supplementary Fig. 
\ref{fig:robustness_and_bias} a).
\medskip

\noindent In a second round of experiments, we tried to understand what biological 
information used by RWR-M was critical for test gene retrieval. To this 
purpose, we started by only randomizing only the PPI network and keeping the original pathways 
network used by RWR-M. In a second step, we did the opposite. We 
discovered that the bias induced by Graphite is more critical in 
predicting test genes than using the original PPI network. 
Indeed, Supplementary Fig. \ref{fig:robustness_and_bias} a) shows that the 
average Recall of the first 200 positions computed on the randomized 
PPI network is always higher than the average Recall@k compute when we 
randomize the (pathways) Graphite network. Finally, we executed RWR-M 
using randomized versions of both networks as inputs to the 
algorithm, and its prediction performance dropped drastically.
\medskip

\noindent Furthermore, to understand if candidate genes predicted using 
randomized PPI networks are more biologically meaningful than those 
returned using the original PPI, we considered the test suite proposed 
by \cite{lazareva2021limits}. We evaluated RWR-M and BRW on the Lung 
Cancer phenotypic condition using the BIOGRID PPI network, both implemented 
in the test suite. Since RWR based algorithms depend on restart 
probability (i.e., greater values of $r$ make the algorithms depend 
less on the topology of the network and, in the extreme case when $r = 
1$, the Markov chain converges to the personalization vector making the 
topology of the network completely useless), we set $r = 0.25$. 
Moreover, we chose $\alpha = 1$ and $\beta = 1$ when 
running BRW; in this way, we isolated the bias induced by 
manually curated ontologies and not by gene expression. We then 
validated each predicted gene set using gseapy KEGG Pathways enrichment 
analysis implemented in the test suite and we compared the outcomes returned 
using the original PPI with those obtained using  its randomized 
counterparts: uniform, scale-free, epected degree, and shuffled. 
\medskip

\noindent Supplementary Fig. \ref{fig:robustness_and_bias} b) shows the comparison 
between candidate genes predicted using a randomized network and the 
original one. The x axis represents all 
randomized methods that are available in the test suite, while the y 
axis represents the negative log scale of p-values calculated by the 
test suite using KEGG enrichment analysis provided by \textbf{gseapy} on 
predefined Lung Cancer pathways.

\subsection{Comparison with \cite{gentili2019biological}}\label{subse:new_vs_old}
BRW is the evolution of a framework we proposed in 
\cite{gentili2019biological} in its embryonic form. Both rely on the hypothesis that 
genes involved in the same disease should be more likely to share biological similarities. 
\cite{gentili2019biological} is different from the current work in significant 
ways. Our initial approach 
integrated only Gene Ontology to define the personalization vector and 
the transition matrix of the random walk, something that was done 
in an ad-hoc way. In contrast, the framework proposed in this paper 
proposes a principled approach, described in Section 
\ref{section:materials_methods} of the main paper and Sections 
\ref{subsubse:int_pv} and \ref{subse:mat_agg} of this supplementary 
material, whereby multiple, heterogeneous biological datasets such as 
KEGG, Reactome, and GO, can be seamlessly integrated to define a biased 
random walk with restart. In particular,  this new version effectively
integrates differential expression and co-expression data to improve 
the biological meaning of candidate genes prioritized in the top K 
positions. To demonstrate the power of incorporating multi-omics data 
(i.e., Proteome and Transcriptome) and several manually curated 
ontologies (i.e., KEGG, Reactome, and GO), in this section, we compare 
our framework with the algorithm proposed in 
\cite{gentili2019biological}. In the paragraphs that follow, the 
latter is referred to as BRW CIBCB 2019.

\begin{figure}[t]
\centering
\includegraphics[width=\textwidth]{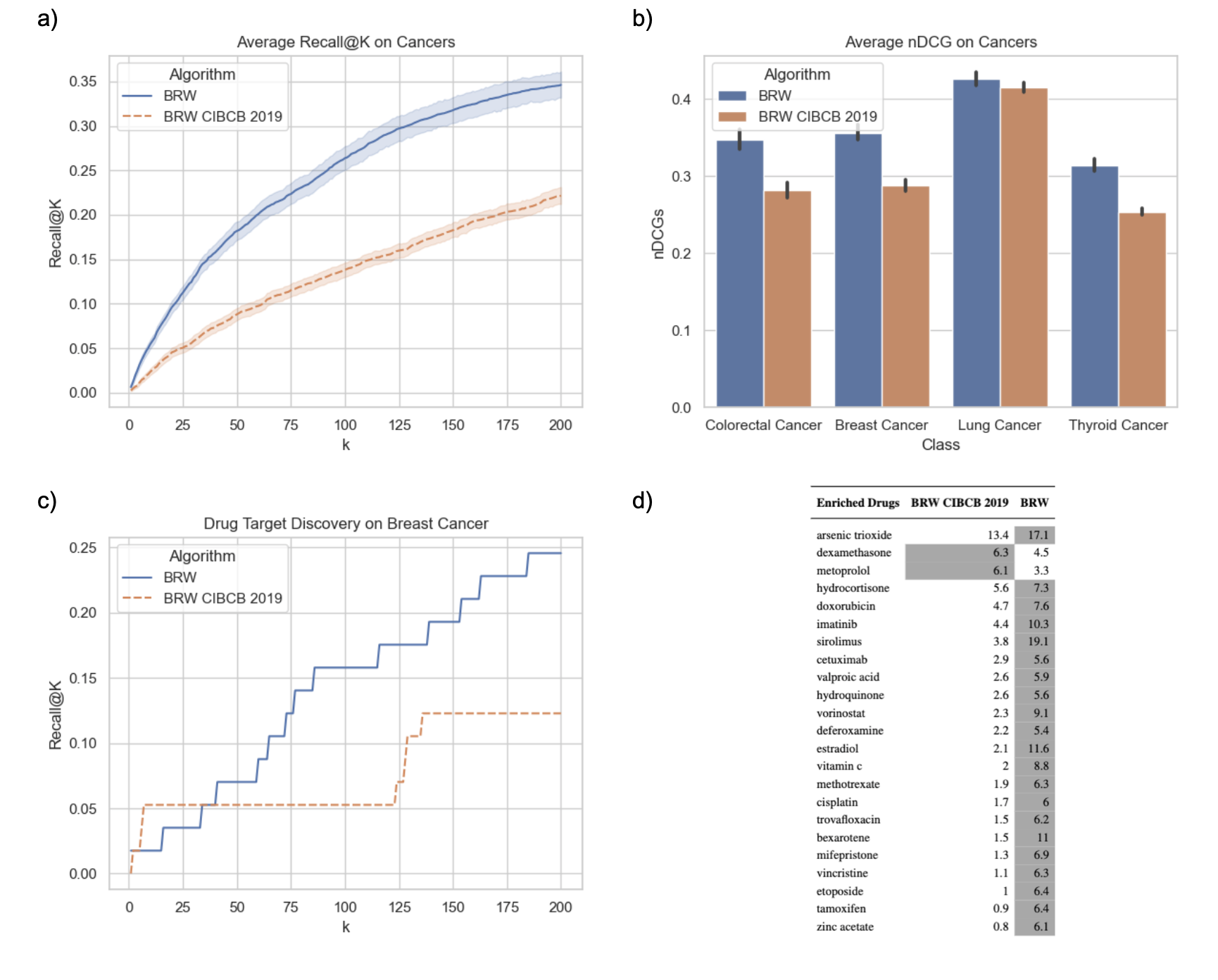}
\caption{Old vs New Approach comparison: a) and b) compare respectively the Recall@K and nDCG of the previous version of BRW with the one presented in this manuscript. c) compares BRW CIBCB 2019 with BRW on the percentage of drug target prioritized in the top 200 positions. Finally, d) compares the drug enrichment computed using the first 200 candidate genes prioritized by the frameworks.}
\label{fig:old_vs_new}
\end{figure}

We compared our approach with \cite{gentili2019biological} along two 
axes, similarly to what we did for other baselines: i) we performed a 
100-fold Montecarlo random sample (internal) validation,
dividing the set of known disease genes into a seed set and a test set, 
which we used to compute Recall@K and nDCG; ii) we further performed an 
external validation, to this purpose focusing on Breast 
Cancer and comparing the two competing approaches in terms of the set of statistically significant drugs 
they retrieve if we restrict to candidate genes ranked the first 200 positions. 

Supplementary Figures \ref{fig:old_vs_new} \textbf{a)} and \textbf{b)} 
compare our approach with BRW CIBCB 2019 in terms of average 
Recall@K and nDCG computed over the aforemention 100-fold Montecarlo 
runs. BRW's ability to leverage manually 
curated information from KEGG and Reactome and gene expression data 
(i.e., differential expression and co-expression network) drastically 
improves the algorithm's performance in prioritizing known candidate 
genes from the test set. 

In the case of drug targets, the fraction of drug targets discovered by 
BRW consistently increases with the number $k$ of top ranked genes we 
consider, showing a two-fold improvement over 
\cite{gentili2019biological} (Figure \ref{fig:old_vs_new} 
\textbf{c)}). This behaviour is further confirmed if one considers 
Breast Cancer drugs that are enriched by genes in top 200 positions. In 
particular, Figure \ref{fig:old_vs_new} \textbf{d)} shows that 
BRW returns a list of candidates enriched in doxorubicin, methotrexate 
and tamoxifen, all well-known drugs to treat Breast Cancer 
(https://www.cancer.gov/about-cancer/treatment/drugs/breast).

\begin{table}[!htp]\centering
\caption{Drugs approved to treat Breast Cancer which have at least one drug target predicted by at least one framework. If an algorithm prioritize a target of a specific drug, its cell is highlighted and it contains 1. The combination of Hyper-parameters used by BRW is $\alpha = 0.75$, $\beta = 0.75$, $r = 0.25$ }\label{tab:drug_discovered_table}
\resizebox{\textwidth}{!}{%
\begin{tabular}{lrrrrrrrr}
\textbf{Drug Bank ID} &\textbf{Drug Name} &\textbf{Associated Condition} &\textbf{BRW} &\textbf{DADA} &\textbf{DIAMOnD} &\textbf{RWR-M} &\textbf{RWR} \\
\textbf{DB12001} &Abemaciclib &Metastatic HR + HER2 - breast cancer &\cellcolor[HTML]{A8A8A8}1 &0 &0 &0 &0 \\
\textbf{DB00675} &Tamoxifen &Breast Cancer &\cellcolor[HTML]{A8A8A8}1 &\cellcolor[HTML]{A8A8A8}1 &\cellcolor[HTML]{A8A8A8}1 &\cellcolor[HTML]{A8A8A8}1 &\cellcolor[HTML]{A8A8A8}1 \\
\textbf{DB11828} &Neratinib &Breast Cancer &\cellcolor[HTML]{A8A8A8}1 &\cellcolor[HTML]{A8A8A8}1 &0 &\cellcolor[HTML]{A8A8A8}1 &\cellcolor[HTML]{A8A8A8}1 \\
\textbf{DB01259} &Lapatinib &Metastatic Breast Cancer &\cellcolor[HTML]{A8A8A8}1 &\cellcolor[HTML]{A8A8A8}1 &0 &\cellcolor[HTML]{A8A8A8}1 &\cellcolor[HTML]{A8A8A8}1 \\
\textbf{DB09073} &Palbociclib &Metastatic Breast Cancer &\cellcolor[HTML]{A8A8A8}1 &0 &0 &0 &0 \\
\textbf{DB11730} &Ribociclib &Metastatic Breast Cancer &\cellcolor[HTML]{A8A8A8}1 &0 &0 &0 &0 \\
\textbf{DB00072} &Trastuzumab &Breast Cancer &\cellcolor[HTML]{A8A8A8}1 &\cellcolor[HTML]{A8A8A8}1 &0 &\cellcolor[HTML]{A8A8A8}1 &\cellcolor[HTML]{A8A8A8}1 \\
\textbf{DB00570} &Vinblastine &Refractory Breast cancer &\cellcolor[HTML]{A8A8A8}1 &\cellcolor[HTML]{A8A8A8}1 &\cellcolor[HTML]{A8A8A8}1 &\cellcolor[HTML]{A8A8A8}1 &0 \\
\textbf{DB11760} &Talazoparib &Metastatic Breast Cancer &\cellcolor[HTML]{A8A8A8}1 &0 &\cellcolor[HTML]{A8A8A8}1 &0 &0 \\
\textbf{DB09074} &Olaparib &Metastatic Breast Cancer &\cellcolor[HTML]{A8A8A8}1 &0 &\cellcolor[HTML]{A8A8A8}1 &0 &0 \\
\textbf{DB00624} &Testosterone &Inoperable, metastatic Breast cancer &\cellcolor[HTML]{A8A8A8}1 &\cellcolor[HTML]{A8A8A8}1 &\cellcolor[HTML]{A8A8A8}1 &\cellcolor[HTML]{A8A8A8}1 &\cellcolor[HTML]{A8A8A8}1 \\
\textbf{DB06710} &Methyltestosterone &Metastatic Breast Cancer &\cellcolor[HTML]{A8A8A8}1 &\cellcolor[HTML]{A8A8A8}1 &\cellcolor[HTML]{A8A8A8}1 &\cellcolor[HTML]{A8A8A8}1 &\cellcolor[HTML]{A8A8A8}1 \\
\textbf{DB01248} &Docetaxel &Metastatic Breast Cancer &\cellcolor[HTML]{A8A8A8}1 &\cellcolor[HTML]{A8A8A8}1 &0 &0 &\cellcolor[HTML]{A8A8A8}1 \\
\textbf{DB01248} &Docetaxel &Locally Advanced Breast Cancer (LABC) &\cellcolor[HTML]{A8A8A8}1 &\cellcolor[HTML]{A8A8A8}1 &0 &0 &\cellcolor[HTML]{A8A8A8}1 \\
\textbf{DB08871} &Eribulin &Refractory, metastatic Breast cancer &\cellcolor[HTML]{A8A8A8}1 &\cellcolor[HTML]{A8A8A8}1 &0 &0 &\cellcolor[HTML]{A8A8A8}1 \\
\textbf{DB00286} &Conjugated estrogens &Metastatic Breast Cancer &\cellcolor[HTML]{A8A8A8}1 &\cellcolor[HTML]{A8A8A8}1 &\cellcolor[HTML]{A8A8A8}1 &0 &\cellcolor[HTML]{A8A8A8}1 \\
\textbf{DB00481} &Raloxifene &Invasive Breast Cancer &\cellcolor[HTML]{A8A8A8}1 &\cellcolor[HTML]{A8A8A8}1 &\cellcolor[HTML]{A8A8A8}1 &0 &\cellcolor[HTML]{A8A8A8}1 \\
\textbf{DB00445} &Epirubicin &Breast Cancer &0 &0 &\cellcolor[HTML]{A8A8A8}1 &0 &0 \\
\textbf{DB01204} &Mitoxantrone &Metastatic Breast Cancer &0 &0 &\cellcolor[HTML]{A8A8A8}1 &0 &0 \\
\textbf{DB00997} &Doxorubicin &Metastatic Breast Cancer &0 &0 &\cellcolor[HTML]{A8A8A8}1 &0 &0 \\
\end{tabular}

}
\end{table}

\end{appendices}

\end{document}